\documentclass[fleqn,usenatbib]{mnras}

\usepackage{newtxtext,newtxmath}

\usepackage[T1]{fontenc}

\DeclareRobustCommand{\VAN}[3]{#2}
\let\VANthebibliography\thebibliography
\def\thebibliography{\DeclareRobustCommand{\VAN}[3]{##3}\VANthebibliography}

\usepackage{graphicx}	
\usepackage{amsmath}	

\usepackage{amssymb}	
\usepackage{siunitx} 
\usepackage{booktabs}
\usepackage[flushleft]{threeparttable}
\usepackage{float}
\usepackage{cleveref}
\usepackage{pifont}
\usepackage{hyperref}
\usepackage{caption}

\newcommand{\PHOEBE}{\texttt{PHOEBE}}
\newcommand{\Gaia}{\textit{Gaia}}
\newcommand{\TESS}{\textit{TESS}}
\newcommand{\WISE}{\textit{WISE}}
\newcommand{\Halpha}{H$\alpha$}
\newcommand{\cmark}{\ding{51}}%
\newcommand{\xmark}{\ding{55}}%

\newcommand{\SEDTeff}[1]{\ifthenelse{\equal{#1}{J1208}}{4700}{\ifthenelse{\equal{#1}{J1721}}{5100}{\PackageError{SEDTeff}{Invalid argument: #1}{The argument provided to \protect\SEDTeff\space must be either J1208 or J1721.}}}}
\newcommand{\SEDTefferr}[1]{\ifthenelse{\equal{#1}{J1208}}{100}{\ifthenelse{\equal{#1}{J1721}}{20}{\PackageError{SEDTefferr}{Invalid argument: #1}{The argument provided to \protect\SEDTefferr\space must be either J1208 or J1721.}}}}
\newcommand{\SEDRad}[1]{\ifthenelse{\equal{#1}{J1208}}{0.73}{\ifthenelse{\equal{#1}{J1721}}{0.877}{\PackageError{SEDRad}{Invalid argument: #1}{The argument provided to \protect\SEDRad\space must be either J1208 or J1721.}}}}
\newcommand{\SEDRaderr}[1]{\ifthenelse{\equal{#1}{J1208}}{0.02}{\ifthenelse{\equal{#1}{J1721}}{0.008}{\PackageError{SEDRaderr}{Invalid argument: #1}{The argument provided to \protect\SEDRaderr\space must be either J1208 or J1721.}}}}
\newcommand{\AnalyticSpotTeff}[1]{\ifthenelse{\equal{#1}{J1208}}{3,545}{\ifthenelse{\equal{#1}{J1721}}{3,707}{\PackageError{AnalyticSpotTeff}{Invalid argument: #1}{The argument provided to \protect\AnalyticSpotTeff\space must be either J1208 or J1721.}}}}
\newcommand{\AnalyticSpotTeffRatio}[1]{\ifthenelse{\equal{#1}{J1208}}{0.75}{\ifthenelse{\equal{#1}{J1721}}{0.73}{\PackageError{AnalyticSpotTeffRatio}{Invalid argument: #1}{The argument provided to \protect\AnalyticSpotTeffRatio\space must be either J1208 or J1721.}}}}
\newcommand{\FUVexcessSigma}[1]{\ifthenelse{\equal{#1}{J1208}}{4.0}{\ifthenelse{\equal{#1}{J1721}}{3.0}{\PackageError{FUVexcessSigma}{Invalid argument: #1}{The argument provided to \protect\FUVexcessSigma\space must be either J1208 or J1721.}}}}
\newcommand{\WIVexcessSigma}[1]{\ifthenelse{\equal{#1}{J1208}}{1.0}{\ifthenelse{\equal{#1}{J1721}}{0}{\PackageError{WIVexcessSigma}{Invalid argument: #1}{The argument provided to \protect\WIVexcessSigma\space must be either J1208 or J1721.}}}}
\newcommand{\BinaryMassFunction}[1]{\ifthenelse{\equal{#1}{J1208}}{0.20}{\ifthenelse{\equal{#1}{J1721}}{0.30}{\PackageError{BinaryMassFunction}{Invalid argument: #1}{The argument provided to \protect\BinaryMassFunction\space must be either J1208 or J1721.}}}}
\newcommand{\PrimaryMass}[1]{\ifthenelse{\equal{#1}{J1208}}{$0.71^{+0.07}_{-0.06}$}{\ifthenelse{\equal{#1}{J1721}}{$0.88^{+0.10}_{-0.08}$}{\PackageError{PrimaryMass}{Invalid argument: #1}{The argument provided to \protect\PrimaryMass\space must be either J1208 or J1721.}}}}
\newcommand{\nDiscTargets}{18}

\title[MS-WD Binaries]{A hidden population of massive white dwarfs: two spotted K+WD binaries}

\author[D. M. Rowan et al.]{D. M.
Rowan,$^{1,2,*}$\thanks{E-mail: rowan.90@osu.edu},
T. Jayasinghe$^{3,*,\dagger}$,
M. A. Tucker$^{1,2,\ddagger}$,
C. Y. Lam$^{3}$,
Todd A. Thompson$^{1,2,4}$,
\newauthor
C. S. Kochanek$^{1,2}$,
N. S. Abrams$^{3}$,
B. J. Fulton$^{5}$,
I. Ilyin$^{6}$,
H. Isaacson$^{3,7}$,
J. Lu$^{3}$,
D. V. Martin$^{1,2}$,
\newauthor
B. Nicholson$^{7,8}$
\\
$^{1}$Department of Astronomy, The Ohio State University, 140 West 18th Avenue, Columbus, OH, 43210, USA\\
$^{2}$Center for Cosmology and Astroparticle Physics, The Ohio State University, 191 W. Woodruff Avenue, Columbus, OH, 43210, USA\\
$^{3}$Department of Astronomy,  University of California Berkeley, Berkeley CA 94720, USA\\
$^{4}$Department of Physics, The Ohio State University, Columbus, Ohio, 43210, USA\\
$^{5}$NASA Exoplanet Science Institute/Caltech-IPAC, Pasadena, CA 91125, USA\\
$^{6}$Leibniz Institute for Astrophysics Potsdam (AIP), An der Sternwarte 16,
D-14482 Potsdam, Germany\\
$^{7}$University of Southern Queensland, Centre for Astrophysics, Toowoomba 4350, Australia\\
$^{8}$Sub-Department of Astrophysics, University of Oxford, Keble Rd, Oxford OX13RH, UK\\
$^{*}$These authors contributed equally to this work.\\
$^{\dagger}$NASA Hubble Fellow\\
$^{\ddagger}$CCAPP Fellow
}

\date{Accepted XXX. Received YYY; in original form ZZZ}

\pubyear{2023}

\begin{document}
\label{firstpage}
\pagerange{\pageref{firstpage}--\pageref{lastpage}}
\maketitle

\begin{abstract}

The identification and characterization of massive ($\gtrsim 0.8 ~M_\odot$) white dwarfs is challenging in part due to their low luminosity. Here we present two candidate single-lined spectroscopic binaries, \Gaia{} DR3 4014708864481651840 and 5811237403155163520, with K-dwarf primaries and optically dark companions. Both have orbital periods of $P\sim 0.45$~days and show rotational variability, ellipsoidal modulations, and high-amplitude radial velocity variations. Using light curves from the \textit{Transiting Exoplanet Survey Satellite} (\textit{TESS}), radial velocities from ground-based spectrographs, and spectral energy distributions, we characterize these binaries to describe the nature of the unseen companion. We find that both systems are consistent with a massive white dwarf companion. Unlike simple ellipsoidal variables, star spots cause the light curve morphology to change between \TESS{} sectors. We attempt to constrain the orbital inclination using \PHOEBE{} binary light curve models, but degeneracies in the light curves of spotted stars prevent a precise determination. Finally, we search for similar objects using \Gaia{} DR3 and \TESS{}, and comment on these systems in the context of recently claimed compact object binaries. 

\end{abstract}




\section{Introduction}

Close binary systems that go through common envelope (CE) evolution can produce a number of unique astrophysical phenomena such as Type Ia supernovae progenitors, cataclysmic variables, and X-ray binaries \citep[e.g.,][]{Paczynski76, Webbink84}. Modeling CE evolution is challenging due to the short timescales and the combination of physical processes involved \citep[e.g.,][]{Ivanova13, Ropke23}. Standard prescriptions such as the energy formalism, which parameterizes how the dissipated orbital energy is used to eject the envelope \citep{Webbink84}, are used for individual binaries \citep[e.g.,][]{Afsar08}, simulations \citep[e.g.,][]{Sandquist98}, and in binary population synthesis \citep[e.g.,][]{Politano10}. However, these model parameters are expected to be time-dependent and vary with stellar properties, which makes producing a predictive model using this formalism challenging \citep{DeMarco11, Ropke23}. By observing the products of CE evolution, we can improve our understanding of binary evolution pathways and the mass distribution of stellar remnants.

In recent years there has been great interest in searching for non-interacting binaries that contain a compact object \citep[e.g.,][]{Breivik17}. While these searches typically focus on identifying black hole binaries \citep{ElBadry23, Chakrabarti23, Tanikawa23}, the same astrometric \citep[e.g.,][]{Andrews19}, spectroscopic \citep[e.g.,][]{Jayasinghe23}, and photometric tools \citep[e.g.,][]{Rowan21, Green23} have been applied to identify neutron star candidates \citep{Zheng22b, Lin23}. 

Many of the false-positives in the search for non-interacting black hole binaries are actually luminous binaries, often with deceptive mass transfer histories \citep[e.g.,][]{Jayasinghe22, ElBadry22_zoo}. Massive white dwarfs can also be detected as ``false-positives'' in these surveys, and measuring white dwarf (WD) mass distributions is relevant to understanding the pulsating phases of the asymptotic giant branch as well as the chemical evolution of galaxies \citep{Cummings18, Catalan08}. The WD mass distribution is generally understood to peak at $M\sim 0.6\ M_\odot$ with an additional peak near $M\sim 0.8\ M_\odot$ \citep{Camisassa19}. Detecting and characterizing WDs in the high-mass tail of this distribution is important for understanding the properties of these dense stellar remnants and the progenitors of Type Ia supernovae. Outside of binary systems, detecting isolated massive white dwarfs is challenging since more massive WDs are more compact and therefore are less luminous. However, by observing the radial velocity and photometric variability of luminous companions, we should be able to find many examples of non-interacting massive WDs.

The majority of WD binaries in main sequence binaries have M-dwarf companions \citep{RebassaMansergas10}. Since CE evolution is expected to be dependent on the mass of both stars, there have been efforts to identify WD+FGK binaries using broadband photometry and UV spectroscopy \citep{Parsons15, Hernandez21, Hernandez22a}, but few massive WDs have been identified through this approach \citep[e.g.,][]{Wonnacott93, Hernandez22b}.

Here, we present two candidate post common envelope white dwarfs with K-dwarf companions identified through radial velocity observations. In Section \S\ref{sec:targets}, we describe how these systems were identified and the follow up radial velocity observations. In Section \S\ref{sec:characterization}, we combine the RVs with broad-band photometry to characterize the binaries and their photometric variability. The late-type main sequence stars are chromospherically active in both binaries, producing star spots that modify the observed ellipsoidal variability. In Section \S\ref{sec:lcs}, we show the limitations that star spots place on our ability to constrain the white dwarf mass. Finally, in Section \S\ref{sec:discussion}, we describe these systems in context with other white dwarf and neutron star binaries detected in radial velocity surveys. 

\section{Target Identification \& Observations} \label{sec:targets}

{\renewcommand{\arraystretch}{1.2}
\begin{table}
    \centering
    \begin{threeparttable}
        \caption{Summary information for J1208 and J1721. The orbital periods are determined from the \TESS{} light curves (Section \S\ref{sec:photometric_variability}) and the velocity semi-amplitudes are measured from the spectroscopic orbits (Section \S\ref{sec:rv_orbits}). $N_{\rm{TESS}}$ and $N_{\rm{RV}}$ report the number of \TESS{} sectors and the number of RV observations, respectively. We use spectral energy distributions to estimate the photometric primary mass and radius (Section \S\ref{sec:seds}). Extinctions are estimated using \texttt{mwdust} \citep{Bovy16}.} \label{tab:summary_table}
        \begin{tabular}{lll}
\toprule
{} &                   J1208 &                   J1721 \\
\midrule
RA $(^\circ)$              &    $182.01140592783395$ &     $260.4612142570771$ \\
DEC $(^\circ)$             &     $31.18433176750879$ &    $-68.74176836793418$ \\
GDR3 Source                &     4014708864481651840 &     5811237403155163520 \\
Distance (pc)              &    $88.6^{+0.1}_{-0.2}$ &   $250.5^{+0.6}_{-0.8}$ \\
$A_{\rm{RV}}\ (\rm{km/s})$ &                     366 &                     959 \\
Gaia $G$ (mag)             &                   11.42 &                   12.68 \\
$N_{\rm{TESS}}$            &                       2 &                       3 \\
Period (d)                 &     $0.46319\pm0.00004$ &     $0.44690\pm0.00003$ \\
$N_{\rm{RV}}$              &             12$\dagger$ &                       5 \\
$K\ (\rm{km/s})$           &               $161\pm2$ &               $186\pm3$ \\
$f(M)\ (M_\odot)$          &                    0.20 &                    0.30 \\
$M_1\ (M_\odot)$           &  $0.71^{+0.07}_{-0.06}$ &  $0.88^{+0.10}_{-0.08}$ \\
$R_1\ (R_\odot)$           &           $0.73\pm0.02$ &         $0.877\pm0.008$ \\
$[$M/H$]$                  &                  $-0.2$ &                 $-0.08$ \\
$A_V\ (\rm{mag})$          &                    0.00 &                    0.11 \\
UV Excess                  &                  \cmark &                  \cmark \\
X-ray Detection            &                  \cmark &                  \xmark \\
\bottomrule
\end{tabular}

        \begin{tablenotes}
            \fontsize{6.5}{7.8}
            \item $^{\dagger}$ We exclude the three LAMOST observations from the RV orbit fits since they occur $\gtrsim 4400$ cycles before the PEPSI/APF observations.
        \end{tablenotes}
    \end{threeparttable}
\end{table}}

We identified high-amplitude RV and photometric variability in two K-dwarfs, 
LAMOST J120802.64+311103.9 \citep[hereafter J1208,][]{Cui12} and \Gaia{} DR3 5811237403155163520 \citep[hereafter J1721,][]{Gaia16, Gaia22}. Both systems show short-period photometric variability in \TESS{} \citep{Ricker15} consistent with ellipsoidal variability. For both systems we obtain follow-up radial velocity observations to fully characterize the binary orbits. Table \ref{tab:summary_table} reports summary parameters of these two targets. In Sections \S\ref{sec:targets_j1208} and \S\ref{sec:targets_j1721} we describe how we identified these targets and the spectroscopic and photometric observations used to characterize them. 

\subsection{J1208} \label{sec:targets_j1208}

\begin{figure*}
    \centering
    \includegraphics[width=0.8\linewidth]{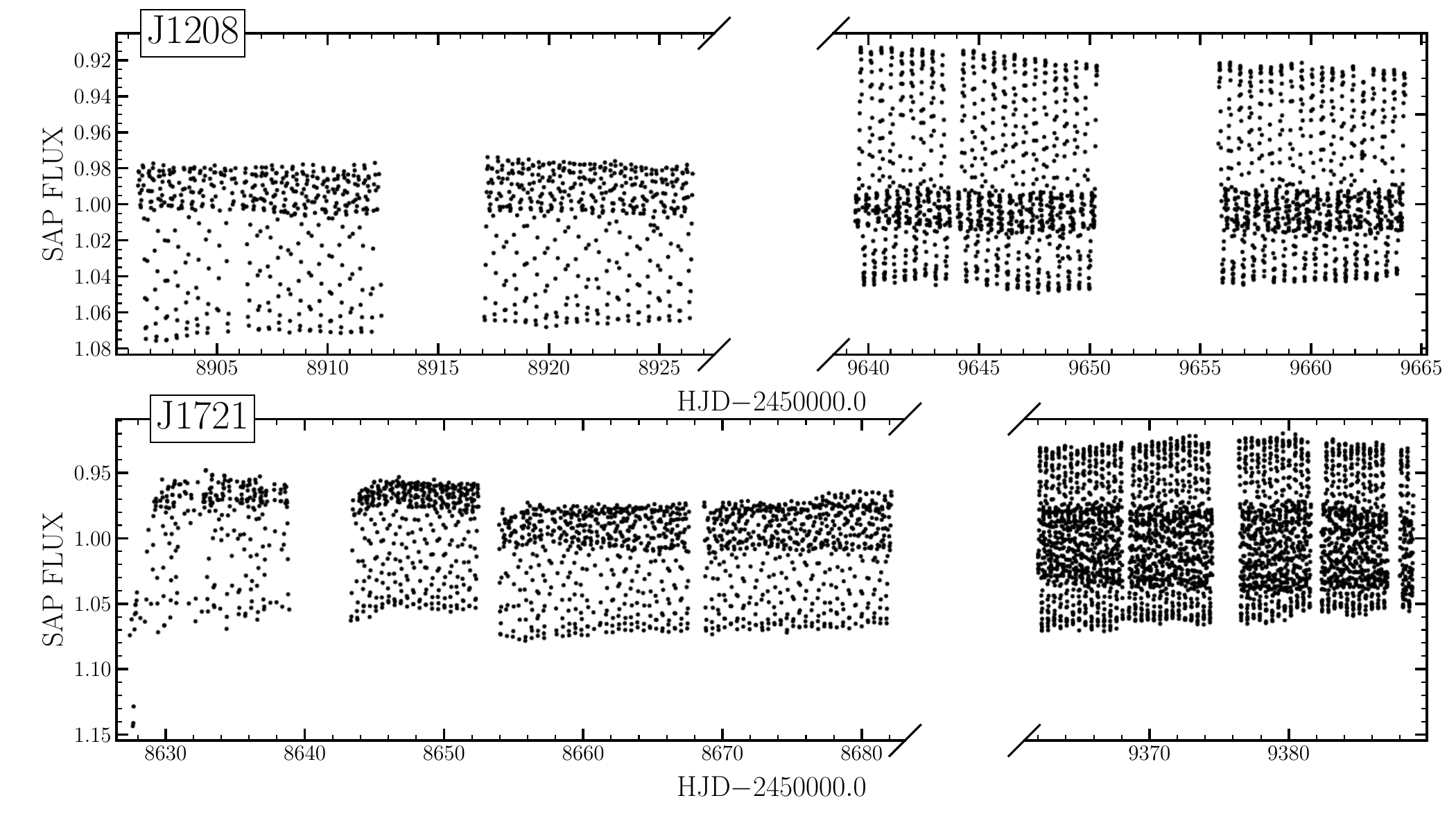}
    \caption{\TESS{} light curves of J1208 (top) and J1721 (bottom). Both targets show clear periodic variability as well as long-term modulations between \TESS{} sectors.}
    \label{fig:tess_lcs_unfolded}
\end{figure*}

\begin{table}
    \centering
    \caption{Radial velocity observations for J1208.}
    \sisetup{table-auto-round,
     group-digits=false}
    \setlength{\tabcolsep}{12pt}
    \begin{center}
        \begin{tabular}{S[table-format=7.5] S[table-format=3.2] S[table-format=1.2] r}
\toprule
         {JD} &    {RV} &  {$\sigma_{\rm{RV}}$} & {Instrument} \\
{} & {(km/s)} & {(km/s)} & {}\\ 
\midrule
2456423.04236 &   43.46 &                  5.36 &   LAMOST-LRS \\
2457094.18750 & -166.40 &                  4.49 &   LAMOST-LRS \\
2457739.42917 & -128.55 &                  6.33 &   LAMOST-LRS \\
2459984.86601 & -127.30 &                  4.40 &        PEPSI \\
2459984.94437 & -171.60 &                  4.40 &        PEPSI \\
2459985.03646 &  -25.40 &                  4.50 &        PEPSI \\
2459989.01660 & -100.67 &                  3.78 &          APF \\
2459984.96948 & -140.56 &                  3.65 &          APF \\
2459992.94531 &   60.92 &                  3.89 &          APF \\
2460084.75719 &  153.85 &                  3.78 &          APF \\
2460085.83655 &  -97.89 &                  3.84 &          APF \\
2460111.81262 & -152.04 &                  4.00 &          APF \\
\bottomrule
\end{tabular}

    \end{center}
    \label{tab:j1208_rvs}
\end{table}

J1208 (\Gaia{} DR3 4014708864481651840) was originally identified as a non-interacting compact object binary candidate by \citet{Mu22} using multi-epoch spectra from the Large Sky Area Multi-Object Fiber Spectroscopy Telescope \citep[LAMOST,][]{Cui12}. There are three low-resolution LAMOST spectra of J1208, taken on 2013 May 10, 2015 March 12, and 2016 December 16. \citet{Mu22} report a RV amplitude $\Delta V_R = 262$~km/s and a spectral type of K-dwarf plus dwarf carbon star \citep[K3+dCK,][]{Roulston20}. They identify photometric variability with a period of $P=0.4630$~days and find a binary mass function $f(M)=0.11\ M_\odot$. \Gaia{} also reports a large radial velocity amplitude $A_{\rm{RV}}=366$~km/s for this target. This value is computed as the difference between the largest and smallest RVs measured after outlier removal.

We obtained three additional high-resolution ($R\approx 43,000$) spectra on 2023 February 9 using the Potsdam Echelle Polarimetric and Spectroscopic Instrument \citep[PEPSI,][]{Strassmeier15} on the \textit{Large Binocular Telescope}. Each observation had a 10~min integration time with the 300$\mu$m fiber and two cross-dispersers covering 4758--5416~\AA{} and 6244--7427~\AA{}. We also obtained six observations with the Automated Planet Finder (APF) Levy spectrograph at the Lick Observatory \citep[$R\approx 80,000$,][]{Vogt14} on 2023 February 9, 13, and 17, 2023 May 20 and 21, and 2023 June 16. The first observation had an integration time of 10 min and the others had an integration time of 15 min. The observations used the $2\arcsec\times3\arcsec$ Decker-T slit. The APF spectra have a wavelength range of 3730--10206\AA{} and the raw 2D echelle spectra are reduced to 1D spectra through the California Planet Survey \citep[CPS,][]{Howard10} pipeline. Next, the 1D echelle spectra are continuum normalized and the orders are combined. APF and PEPSI radial velocities were derived by cross-correlating the continuum normalized spectrum with synthetic spectra using iSpec \citep{BlancoCuaresma14, BlancoCuaresma19} with the templates broadened to match the resolution of the data. Table \ref{tab:j1208_rvs} reports the radial velocity observations of J1208. 

J1208 was observed by the \TESS{} in sectors 22 (2020 March) and 49 (2022 March). We downloaded light curves from the Quick-Look Pipeline \citep[QLP,][]{Huang20a, Huang20b}. We use the raw, undetrended light curves rather than the detrended light curves since the detrending procedure can often remove variability on timescales $>0.3$~days \citep{Green23}. Each sector shows clear periodic variability, as shown in the top panel of Figure \ref{fig:tess_lcs_unfolded}. We also retrieve archival photometry from the All-Sky Automated Survey \citep[ASAS,][]{Pojmanski97}, the All-Sky Automated Survey for Supernovae \citep[ASAS-SN,][]{Shappee14, Kochanek17, Hart23} and the Wide-field Infrared Survey Explorer \citep[\WISE{},][]{Wright10}. 

\subsection{J1721} \label{sec:targets_j1721}

\begin{table}
    \centering
    \caption{Radial velocity observations for J1721}
    \sisetup{table-auto-round,
     group-digits=false}
    \setlength{\tabcolsep}{12pt}
    \begin{center}
        \begin{tabular}{S[table-format=7.5] S[table-format=3.2] S[table-format=1.2] r}
\toprule
         {JD} &     {RV} &  {$\sigma_{\rm{RV}}$} & {Instrument} \\
{} & {(km/s)} & {(km/s)} & {}\\ 
\midrule
2460004.85545 &  -92.871 &                 2.433 &       CHIRON \\
2460006.89372 &  -60.458 &                 1.013 &       CHIRON \\
2460007.88454 & -221.646 &                 8.559 &       CHIRON \\
2460008.85706 & -142.257 &                 3.132 &       CHIRON \\
2460010.88317 &   23.665 &                 1.379 &       CHIRON \\
\bottomrule
\end{tabular}

    \end{center}
    \label{tab:j1721_rvs}
\end{table}

\begin{figure}
    \centering
    \includegraphics[width=\linewidth]{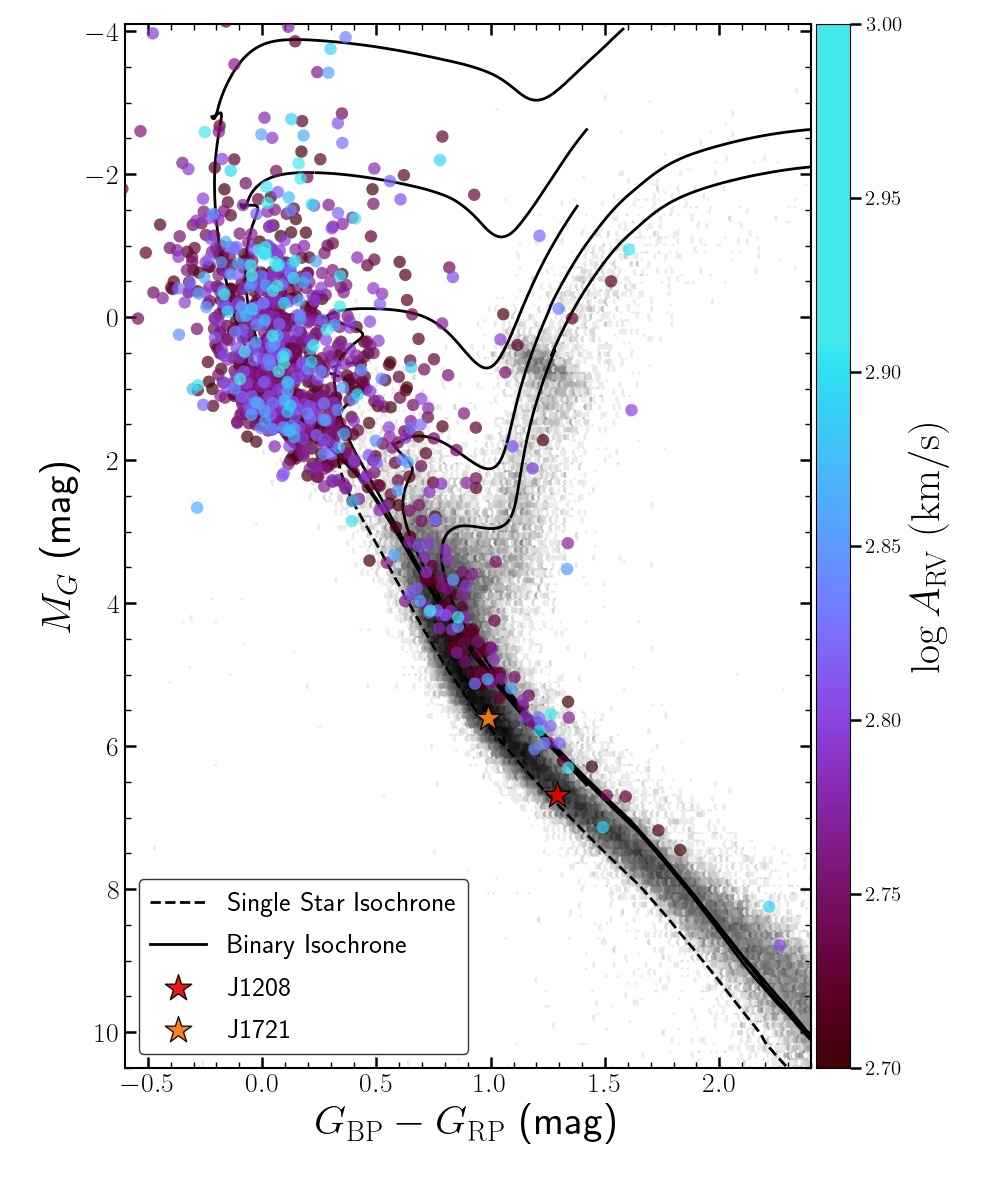}
    \caption{\Gaia{} DR3 color-magnitude diagram (CMD) of a sample of random \textit{Gaia} targets (gray background) and $A_{\rm{RV}} > 500$~km/s targets (colored). The solid black lines show MIST isochrones \citep{Choi16, Dotter16} corresponding to an equal mass binary and the dashed line shows a single-star isochrone. Extinctions are determined using {\tt mwdust} with distances from \citet{BailerJones21}. J1208 and J1721 are shown as the red and orange points.}
    \label{fig:rvamp_cmd}
\end{figure}

\begin{figure*}
    \centering
    \includegraphics[width=\linewidth]{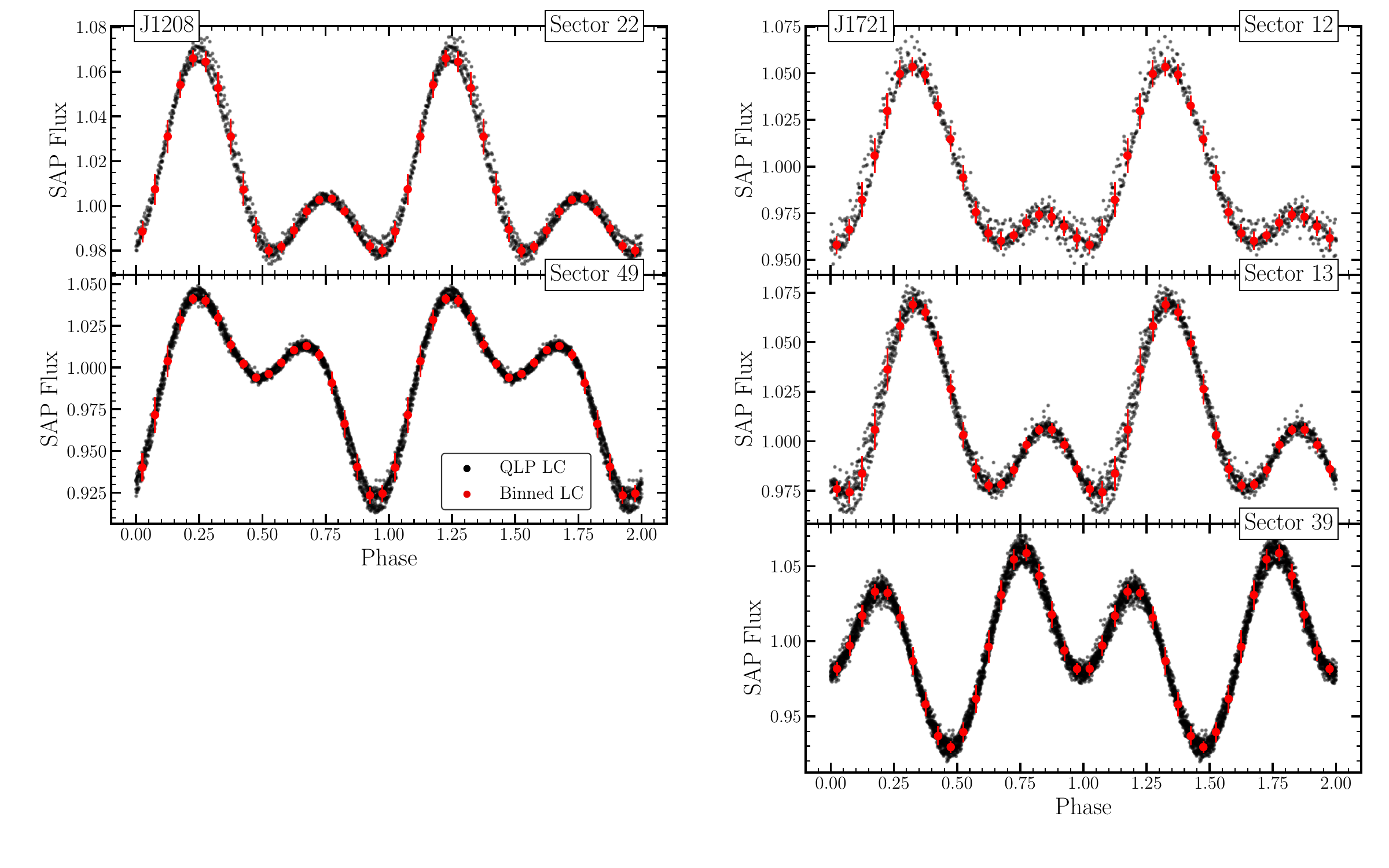}
    \caption{Phase-folded \TESS{} light curves of J1208 (left) and J1721 (right). Both systems show short-period ellipsoidal modulations with asymmetric maxima. The light curve shape varies dramatically between \TESS{} sectors, suggesting the presence of spots that evolve on short timescales. Orbital phase is defined such that RV maxima occurs at phase $\phi=0.75$.}
    \label{fig:tess_lcs_folded}
\end{figure*}

We identified J1721 as a \Gaia{} photometric variable with a high RV amplitude. \Gaia{} characterized this source a short timescale photometric variable with $P=0.22347$~day \citep{Eyer22}. \Gaia{} also reports an \texttt{rv\_amplitude\_robust} of $A_{\rm{RV}} = 959.2$~km/s. For comparison, J1208 has $A_{\rm{RV}} = 366.3$~km/s, and only 1278 stars in \Gaia{} DR3 have $A_{\rm{RV}} > 500.0$~km/s. Figure \ref{fig:rvamp_cmd} shows these 1278 stars on a \Gaia{} color-magnitude diagram (CMD) and highlights J1208 and J1721. The majority of high $A_{\rm{RV}}$ targets sit on the upper main sequence. The narrow wavelength range of the \Gaia{} Radial Velocity Spectrometer \citep[846--870~nm, ][]{Cropper18} was designed to measure radial velocities of cool stars, and RVs for hot stars ($6900<T_{\rm{eff}}<14500$~K) only became available with \Gaia{} DR3 \citep{Blomme22}. It seems likely that many of the high $A_{\rm{RV}}$ stars on the upper main sequence suffer from systematic effects. Below $M_G \lesssim 3.5$~mag, almost all of the high $A_{\rm{RV}}$ targets appear consistent with the binary star main sequence. We selected J1721 for additional follow-up because of its CMD position near a single star isochrone, its periodic photometric variability, and its high radial velocity amplitude.

We obtained multi-epoch spectra with CHIRON \citep{Tokovinin13} on the SMARTS 1.5m telescope \citep{Schwab12} to validate the orbit and determine the nature of the companion. We obtained five spectra, each with the fiber mode, which uses $4\times4$~pixel binning ($R\approx 28,000$), and a Th-Ar comparison lamp. Four observations had 20~min integration times, and one had a 30~min integration time. RVs were derived using a least-squares deconvolution against a non-rotating synthetic spectral template, as in \citet{Zhou20}. Table \ref{tab:j1721_rvs} reports the radial velocity observations of J1721.

J1721 was observed by \TESS{} in sectors 12 (2019 June), 13 (2019 July) and 39 (2021 June). The bottom panel of Figure \ref{fig:tess_lcs_unfolded} shows the variability, which is clearly periodic but changes between \TESS{} sectors. As with J1208, we also retrieve archival photometry from ASAS-SN, WISE, and \Gaia{}. 

\section{Binary Characterization} \label{sec:characterization}

\subsection{Photometric Variability} \label{sec:photometric_variability}

\begin{figure}
    \centering
    \includegraphics[width=\linewidth]{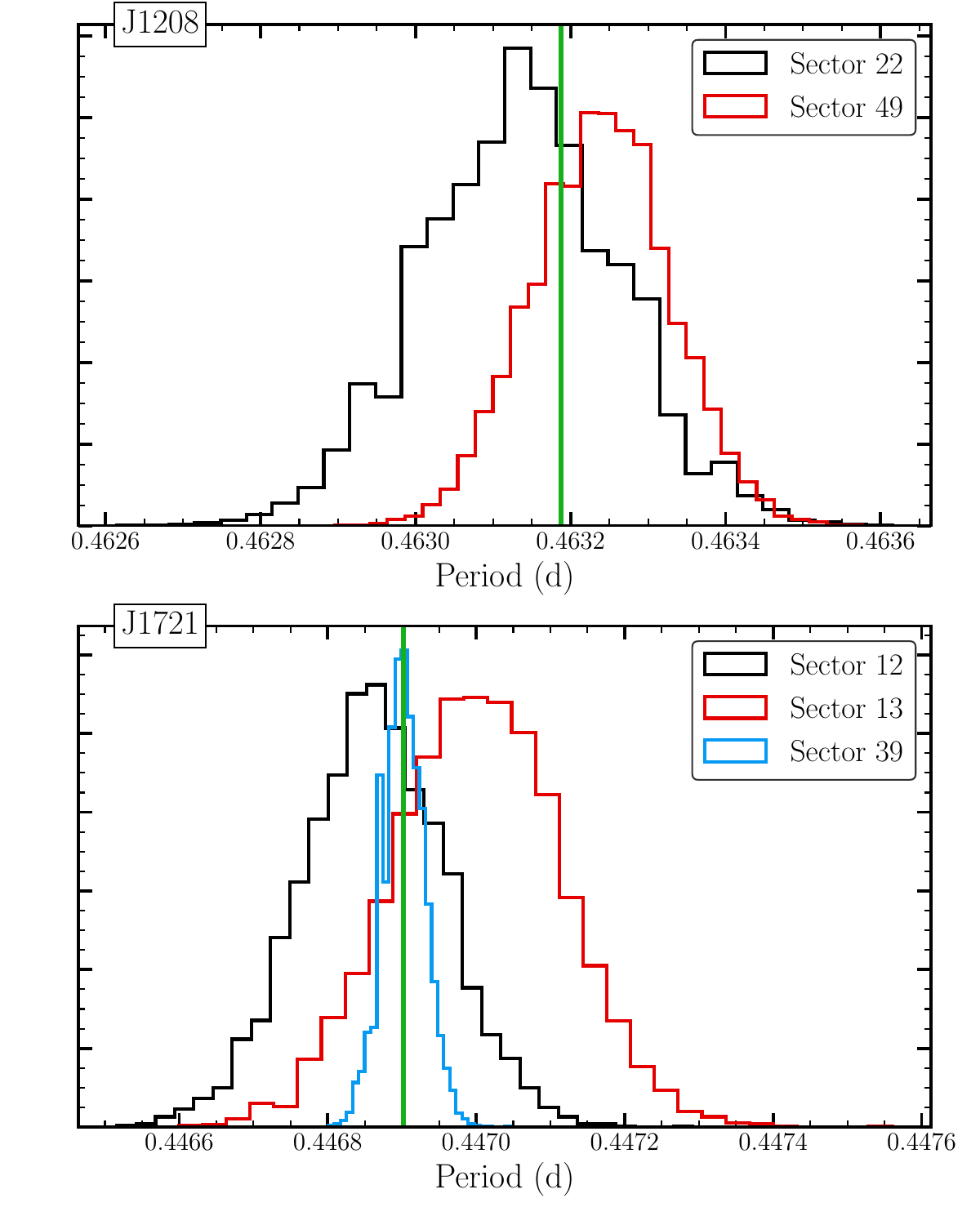}
    \caption{Periods determined from Lomb-Scargle periodograms for each \TESS{} sector for J1208 (top) and J1721 (bottom). The differences in periods for each sector are not statistically significant, and we adopt the median period as the orbital period of the system (vertical green line).}
    \label{fig:period_histograms}
\end{figure}

\begin{figure*}
    \centering
    \includegraphics[width=\linewidth]{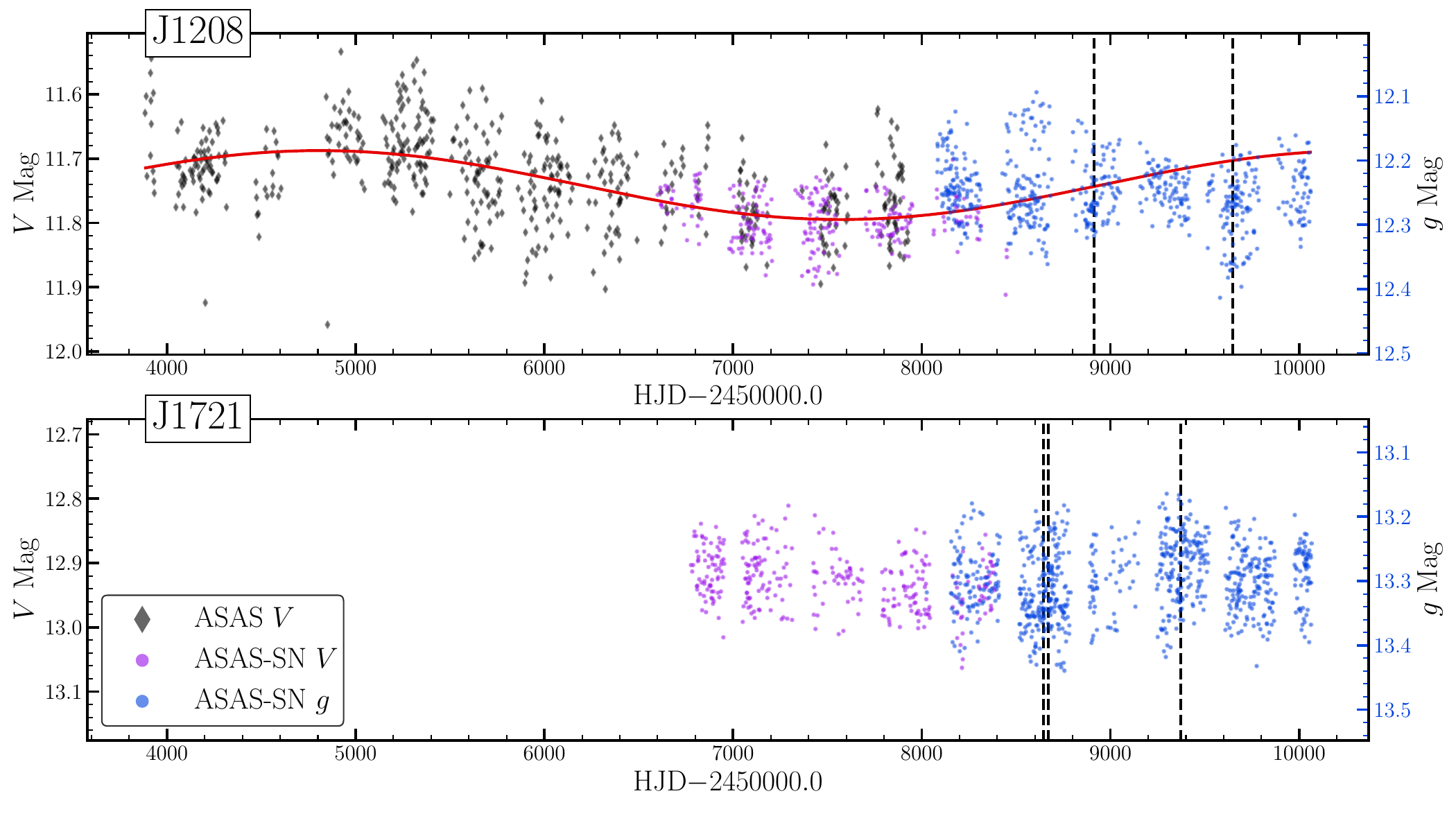}
    \caption{ASAS and ASAS-SN light curves of J1208 (top) and (J1721) bottom. We find some evidence of long-term variability at $\sim 5520$~days in the J1208 light curve, indicated by the red curve. The vertical dashed lines indicate the median times of the \TESS{} sectors shown in Figure \ref{fig:tess_lcs_folded}.}
    \label{fig:archival_unfolded}
\end{figure*}

The \TESS{} light curves shown in Figure \ref{fig:tess_lcs_unfolded} reveal periodic variability in both targets. Since the light curve shape varies between \TESS{} sectors (Figure \ref{fig:tess_lcs_unfolded}), we ran a Lomb-Scargle periodogram \citep{Lomb76, Scargle82} on each sector independently. To estimate the uncertainty in the period, we performed $10^4$ bootstrap iterations for each \TESS{} sector. Figure \ref{fig:tess_lcs_folded} shows the phase-folded light curves for each \TESS{} sector. 

The \TESS{} light curves show ellipsoidal modulations caused by the tidal distortion of the K-dwarf by a close stellar companion. Ellipsoidal variable (ELLs) light curves are typically double-peaked with uneven minima. Both J1208 and J1721 also have uneven maxima in their light curves. While some asymmetry in the maxima is expected in short-period binaries due to relativistic beaming \citep{Loeb03, Masuda19}, the large difference between the light curve maxima and the variations between \TESS{} sectors instead suggests that the K-dwarfs are heavily spotted. Since spots evolve over timescales of tens to hundreds of days \citep[e.g.,][]{Giles17}, the light curve shape changes dramatically between \TESS{} sectors. 

Figure \ref{fig:period_histograms} compares the orbital periods determined for each sector. For each target, the period varies between \TESS{} sectors by $\sim15$ -- $20$~min, but this is not a statistically significant difference. Small variations between the periods of each sector could be evidence of latitudinal differential rotation or slightly asynchronous rotational and orbital periods. Taking the median period from the different sectors, we find $P=0.46319\pm0.00004$ and $P=0.44690\pm0.00003$ for J1208 and J1721, respectively. The period for J1721 is approximately twice the value reported in \Gaia{} DR3 \citep{Eyer22}.

The archival ASAS and ASAS-SN light curves of J1208 shown in Figure \ref{fig:archival_unfolded} suggest long-term variations that could be evidence of spot modulations or star cycles. Both light curves show periodic variability at $\sim 0.46$~days corresponding to the orbital period identified in the \TESS{} light curve. In the combined ASAS+ASAS-SN $V$-band light curve we also find evidence for periodic variability at $\sim 5520$~days ($\sim 15.1$~years), which could be representative of a stellar activity cycle. The ASAS-SN $g$-band data do not appear to follow this trend, and a Lomb-Scargle periodogram of the combined $V$ and $g$-band light curve, with an offset applied to the $g$-band data to align it with the $V$-band data, did not yield significant periods other than the $\sim 0.46$~day signal. There do appear to be long-term variations in the $g$-band data, which is slightly bluer than the $V$-band and includes the calcium H and K lines with rest wavelengths 3969~\AA~and~3934~\AA, respectively. The $V$-band and $g$-band light curves may therefore trace different timescales of stellar activity \citep[e.g.,][]{Mignon23}.

The ASAS-SN light curve of J1721 does not show similar long-term variations despite the clear sector-to-sector variations in the \TESS{} light curve. This could suggest relatively less chromospheric activity, which is consistent with the lack of \Halpha{} emission discussed below (Section \S\ref{sec:halpha}). 

We also inspected the phase-folded light curves from ASAS, ASAS-SN, ATLAS, \Gaia{}, and \WISE{} for both systems when available. Unsurprisingly, the multi-year light curves folded at the periods from the \TESS{} observations have substantial scatter due to spot evolution.

\subsection{Spectroscopic Orbits} \label{sec:rv_orbits}

\begin{figure}
    \centering
    \includegraphics[width=\linewidth]{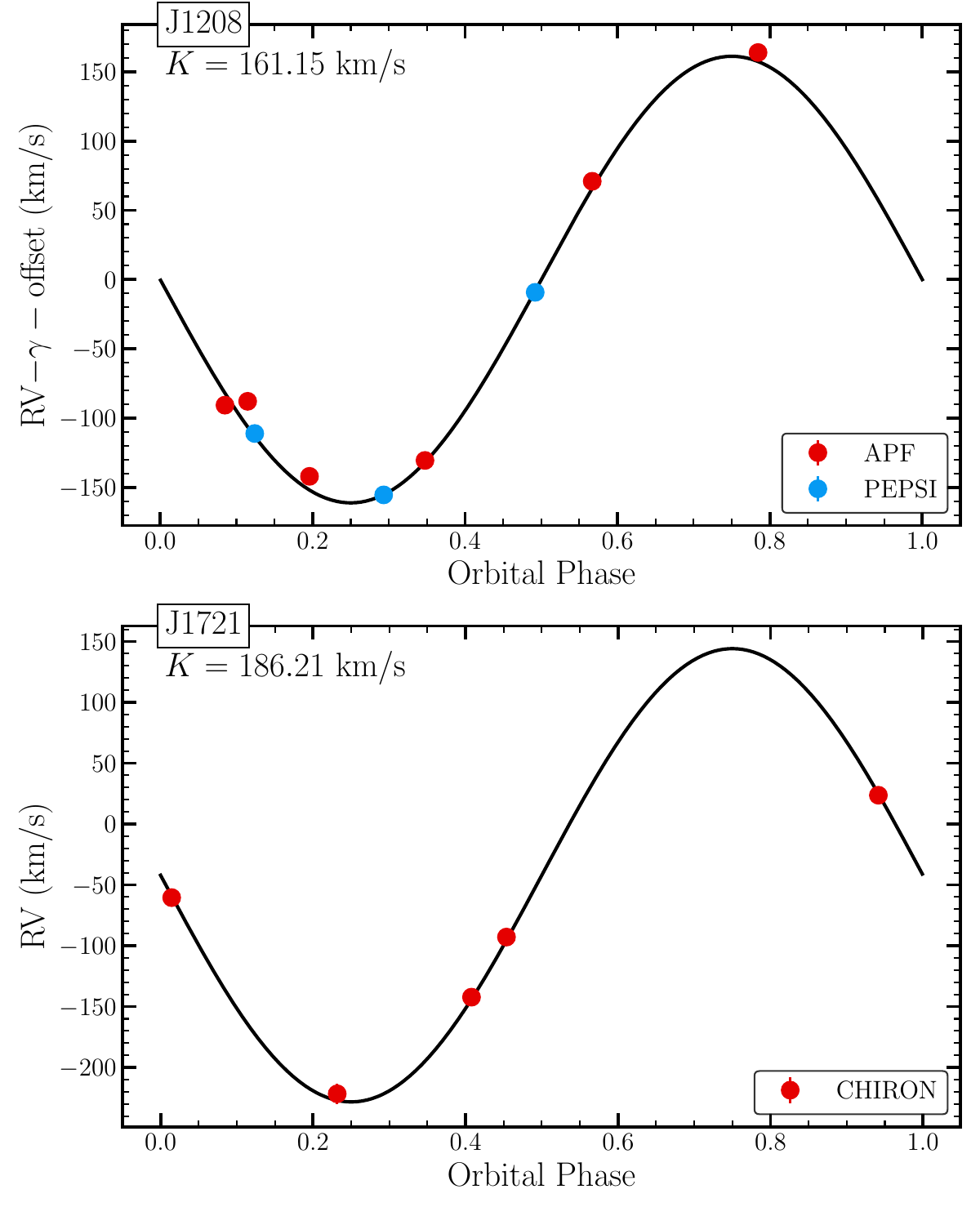}
    \caption{Spectroscopic orbits of J1208 (top) and J1721 (bottom). We fit a circular orbit (Equation \ref{eqn:circular_orbit}) fixed to the photometric period to derive the radial velocity semi-amplitude, $K$. The RV uncertainties are smaller than the point size. For J1208, the radial velocities are shown with the center-of-mass velocity and the PEPSI RV offset subtracted.}
    \label{fig:rv_orbits}
\end{figure}

Since both targets are consistent with short period binaries, we fit a circular Keplarian orbit model of the form
\begin{equation} \label{eqn:circular_orbit}
    \text{RV}(t) = \gamma + K \cos\left(\frac{2\pi}{P}(t-t_0)\right),
\end{equation}
where $K$ is the velocity semi-amplitude, $P$ is the orbital period fixed at the values from the \TESS{} light curves, $t_{0}$ is the time of pericenter passage, and $\gamma$ is the center-of-mass velocity. Since the archival LAMOST observations of J1208 were taken $\gtrsim 4400$ cycles before the PEPSI/APF observations, we chose not to include them in the RV fits since small uncertainties in the orbital period result in large uncertainties in their orbital phase. For J1208, we also fit for an RV offset between the APF and PEPSI measurements. We use the Monte Carlo sampler {\tt emcee} \citep{Foreman-Mackey13} to derive the posteriors on $K$ reported in Table \ref{tab:summary_table}. For both systems, we find that $K < 0.5 A_{\rm{RV}}$, indicating that the \Gaia{} $A_{\rm{RV}}$ is overestimated . For J1208, $K=161\pm2$~km/s is consistent with the $\Delta V_R = 262$~km/s reported by \citep{Mu22} for the three LAMOST observations \citep{Mu22}. Figure \ref{fig:rv_orbits} shows the radial velocity curves where orbital phase is defined such that RV maxima occurs at phase $\phi=0.75$. 

Even though we expect both short-period binaries to be tidally circularized, we also tested models with non-zero eccentricity. The posteriors on the eccentricity for both targets are peaked at zero eccentricity, with an 84th percentile of $e \leq 0.04$ and $e \leq 0.03$ for J1208 and J1721, respectively. 

Assuming a circular orbit, the velocity semi-amplitude, $K$, and orbital period, $P$, are related to the binary masses $M_1$ and $M_2$ and the inclination, $i$, through the binary mass function,
\begin{equation}
    f(M) = \frac{P K^3}{2\pi G} = \frac{M_2^3 \sin^3 i}{(M_1+M_2)^2}.
\end{equation}

\noindent We find $f(M)=\BinaryMassFunction{J1208}\ M_\odot$ for J1208 and $f(M)=\BinaryMassFunction{J1721}\ M_\odot$ for J1721. The binary mass function is the absolute lower-limit on the companion mass $M_2$ obtained for an edge-on binary with the observed star having $M_1=0\ M_\odot$. Additional constraints on the primary mass, $M_1$, and the orbital inclination are then needed to determine the actual companion mass.

\subsection{H$\alpha$ emission} \label{sec:halpha}

\begin{figure}
    \centering
    \includegraphics[width=\linewidth]{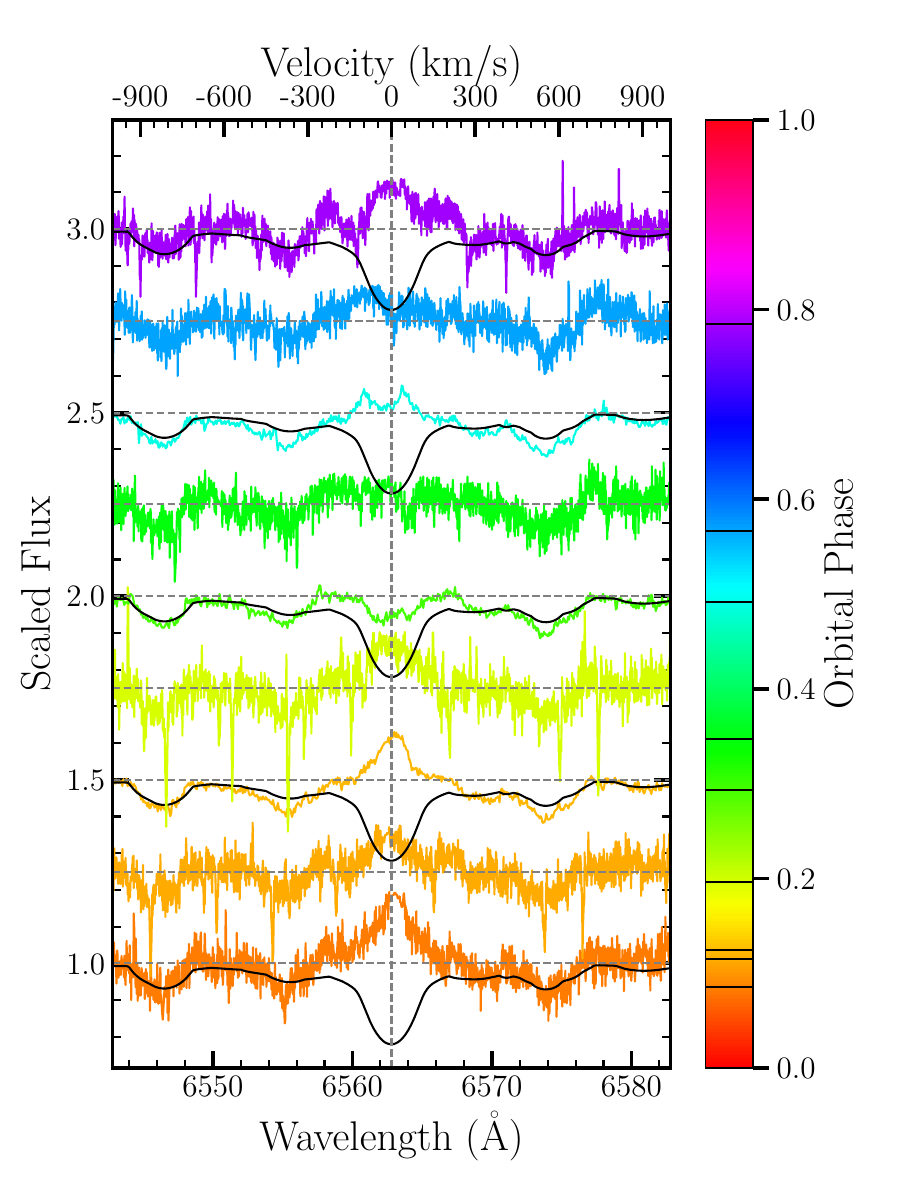}
    \caption{The phase-dependent \Halpha{} emission from J1208 in the  APF and PEPSI spectra. The spectra have been shifted to the rest frame of the K-dwarf. For comparison, in black we show the absorption line from a synthetic spectrum computed using the stellar effective temperature, surface gravity, metallicity, and rotational broadening.}
    \label{fig:j1208_halpha}
\end{figure}

\begin{figure}
    \centering
    \includegraphics[width=\linewidth]{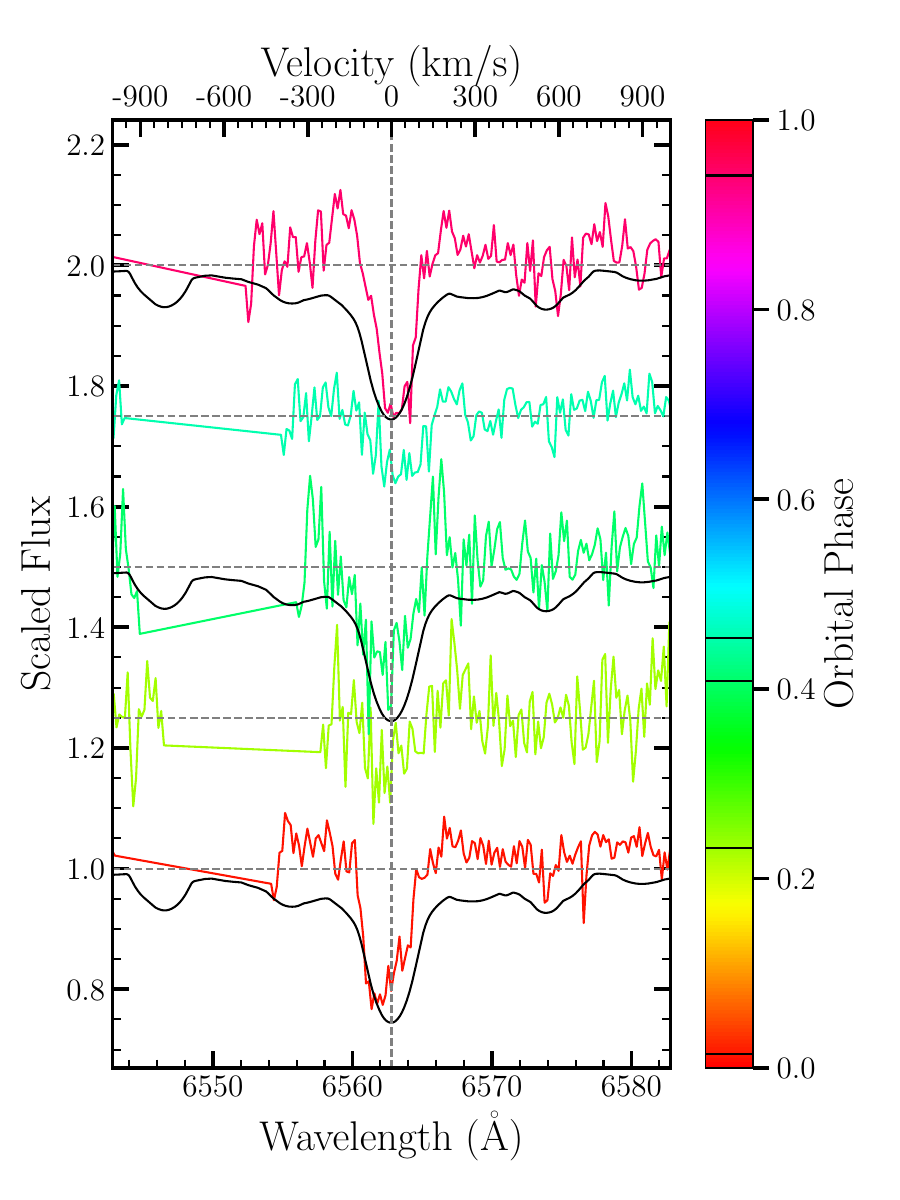}
    \caption{The phase-dependent \Halpha{} absorption feature of J1721 from the CHIRON spectra shifted to the rest frame. Unlike J1208, this binary does not show strong emission features. The spectra have been shifted to the rest frame of the K-dwarf. The black lines show a synthetic spectrum computed using the stellar effective temperature, surface gravity, metallicity, and rotational broadening.}
    \label{fig:j1721_halpha}
\end{figure}

The APF and PEPSI observations of J1028 show variable H$\alpha$ emission (Figure \ref{fig:j1208_halpha}). Orbital phase is defined such that the maximum RV occurs at $\phi=0.75$ and the K star is behind the WD (inferior conjunction) at $\phi=0$. Three observations with orbital phase $0 < \phi < 0.2$ show strong \Halpha{} emission centered on the velocity of the star. One of the two observations between $ 0.3 < \phi < 0.4$ shows \Halpha{} absorption. At $\phi \sim 0.5$, the PEPSI spectra shows double-peaked H$\alpha$ emission. \Halpha{} emission occurs again at $\phi\sim0.8$.

\Halpha{} emission could originate from a combination of chromospheric activity and/or mass transfer. For example, double-peaked emission is commonly seen in accreting compact object binaries \citep[e.g.,][]{Swihart22}. The SED of J1208 also has an apparent excess in the \WISE{} W4 band (Figure \ref{fig:seds}), which could be explained by an accretion disk or dust, but this excess is only significant at the $\WIVexcessSigma{J1208}\sigma$ level. The spectrum with double-peaked emission occurs near phase $0.5$ in J1208, when the K-dwarf is in front of the white dwarf. The \Halpha{} emission appears to track the motion of the K-dwarf primary, which is also consistent with chromospheric emission seen in similar binaries with compact companions \citep[e.g.,][]{Lin23, Zheng22b}.

All the CHIRON spectra of J1721 show \Halpha{} in absorption (Figure \ref{fig:j1721_halpha}). The lack of \Halpha{} emission in J1721 could imply a lower degree of chromospheric activity than in J1208.

The equivalent width of the line changes with orbital phase, with shallower features during phases where the K-dwarf passes in front of the white dwarf. This could suggest that emission from chromospheric activity is filling in the absorption line at these phases. The absorption line is symmetric with respect to the velocity of the K-dwarf, unlike some stripped mass-transfer binaries \citep[e.g., 2MASS J04123153$+$6738486][]{Jayasinghe22, ElBadry22_zoo}. High resolution infrared spectra of the Calcium II triplet (8498, 8542, and 8662\AA) could be used to compare the activity indices of these two targets \citep{Martin17}. The \Gaia{} RVS spectrometer does cover this wavelength range, but neither target has an RVS spectrum included in \Gaia{} DR3. 

\subsection{Spectral Energy Distributions} \label{sec:seds}

\begin{figure}
    \centering
    \includegraphics[width=\linewidth]{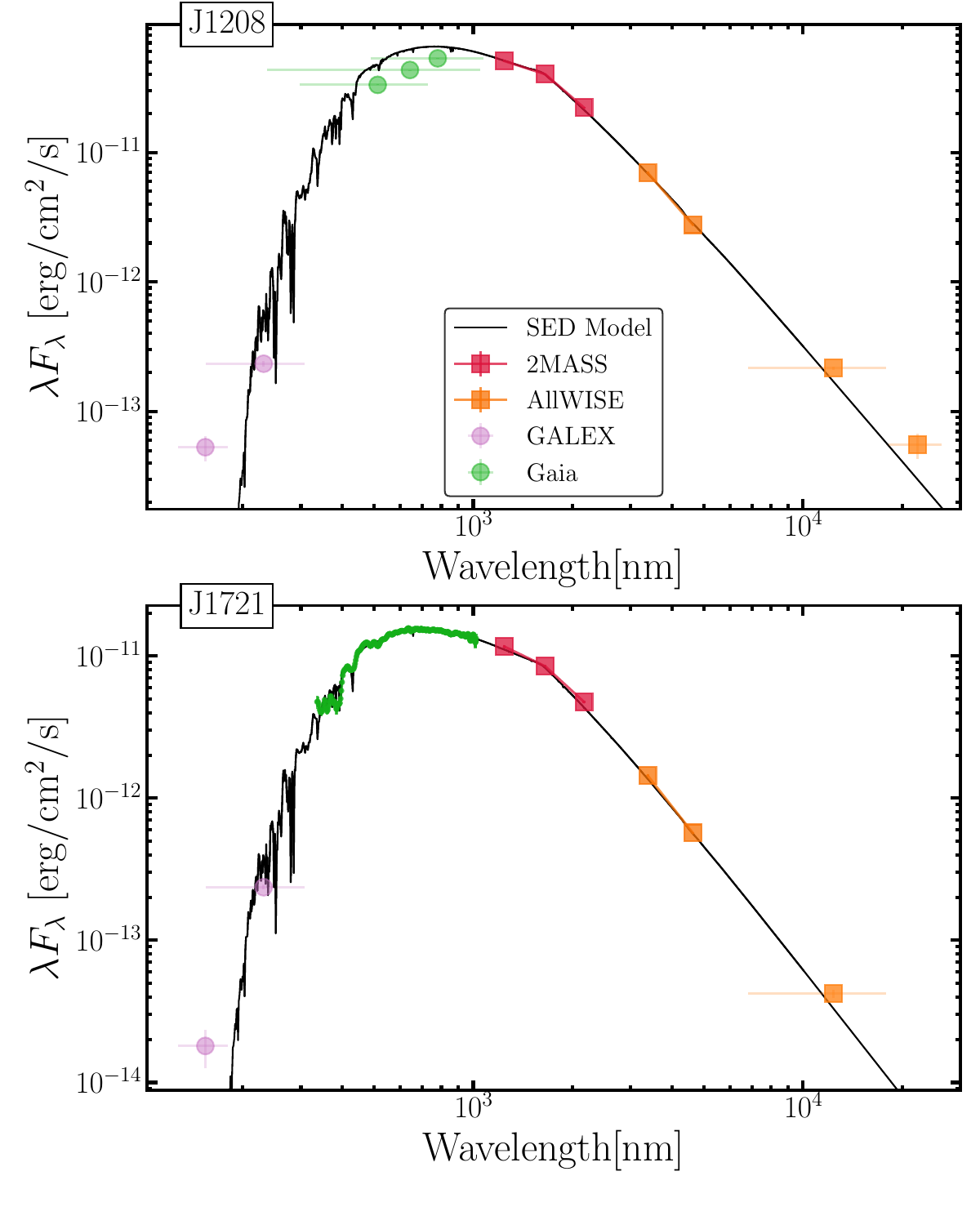}
    \caption{Spectral energy distributions (SEDs) of J1208 (top) and J1721 (bottom). Both are consistent with K-dwarfs and show evidence for a far UV excess in the \textit{GALEX} photometry.}
    \label{fig:seds}
\end{figure}

%

To determine the properites of the K-dwarfs, we start by using broad-band photometry and single-star evolutionary models. We retrieve 2MASS \citep{Cutri03}, \WISE{} \citep{Cutri12}, and  \textit{GALEX} \citep{Bianchi17} photometry for both targets. We also download the low-resolution \Gaia{} XP spectra \citep{DeAngeli22}, which were only available for J1721. For J1208, we use the \Gaia{} $G$, $G_{\rm{BP}}$, and $G_{\rm{RP}}$ magnitudes. We fit the spectral energy distributions (SEDs)
using the \citet{Castelli03} atmosphere models included in {\tt pystellibs}\footnote{\url{https://github.com/mfouesneau/pystellibs}}. We use {\tt pyphot}\footnote{\url{https://mfouesneau.github.io/pyphot/}} to calculate synthetic photometry and sample over stellar parameters with {\tt emcee} \citep{Foreman-Mackey13}. We keep the distance fixed at the values from \citet{BailerJones21} and use $V$-band extinctions from the {\tt mwdust} \citep{Bovy16} ``Combined19'' dust map \citep{Drimmel03, Marshall06,Green19}. We do not include the \textit{GALEX} photometry in our SED fits since the spotted primaries are expected to have additional ultraviolet (UV) flux from chromospheric activity that is not represented in the atmosphere models. WD companions could also contribute to the UV flux. 

Figure \ref{fig:seds} shows the SEDs and the fits. We find that J1208 is consistent with a K-dwarf of radius $R_1=\SEDRad{J1208}\pm\SEDRaderr{J1208}\ R_\odot$ and temperature $T_{\rm{eff}}=\SEDTeff{J1208}\pm\SEDTefferr{J1208}$~K and J1721 has radius $R_1=\SEDRad{J1721}\pm\SEDRaderr{J1721}\ R_\odot$ and temperature $T_{\rm{eff}}=\SEDTeff{J1721}\pm\SEDTefferr{J1721}$~K. We also attempt two-star SED fits and find no acceptable solutions with near equal-mass binaries. While the SED does permit having a low-mass companion ($M_2\lesssim0.5\ M_\odot$), these masses are too small to reproduce the observed radial velocity semi-amplitude, even at edge-on inclinations.

The SEDs of both targets have \textit{GALEX} near-ultraviolet (NUV) magnitudes largely consistent with the K-dwarf model. There is an excess far-ultraviolet (FUV) flux of $>\FUVexcessSigma{J1208}\sigma$ and $>\FUVexcessSigma{J1721}\sigma$ for J1208 and J1721, respectively. The UV excess could be due to chromospheric activity, but, depending on the age of the system, a cool or massive white dwarf could also conceivably produce the observed FUV flux with negligible contributions in the NUV.

J1208 was detected as an X-ray source in ROSAT \citep{Voges99} with a separation of $<15\arcsec$. \citet{Kiraga13} report an X-ray to bolometric flux ratio $\log (F_x/F_b)=-3.25\pm0.25$ for J1208. \citet{Pizzolato03} measured the relationship between ROSAT X-ray flux and rotation period for different mass bins using \textit{Kepler} rotational variables. For stars $0.63 < M/M_\odot < 0.78$, they find that the X-ray to bolometric luminosity ratio saturates at $\log(L_x/L_b) = -3.1\pm0.2$ below rotation periods of $3.3\pm1.5$~days. Since this is consistent with the X-ray luminosity of J1208, the system's X-ray emission could come entirely from chromospheric activity. \textit{SWIFT} UV photometry or Hubble Space Telescope UV spectroscopy could provide meaningful constraints on the nature of this high-energy emission. J1721 has no reported X-ray detection. 

Since both targets appear consistent with a single-star isochrone (Figure \ref{fig:rvamp_cmd}), we can use evolutionary tracks from MESA Isochrones and Stellar Tracks \citep[MIST][]{Dotter16,Choi16} to estimate the mass of the photometric primary. We consider MIST evolutionary tracks covering the mass range 0.5--1.2~$M_\odot$ at the metallicities estimated for each target ($[\rm{Fe/H}]=-0.2$ for J1208 and $[\rm{Fe/H}]=-0.08$ for J1721) from the spectra. For each mass, we construct linear interpolations of the \Gaia{} $G$, $G_{\rm{BP}}$, and $G_{\rm{RP}}$ magnitudes with age. We use {\tt emcee} \citep{Foreman-Mackey13} to sample over primary mass, stellar age, and distance. We use distances from \citet{BailerJones21} and convert $V$-band extinctions from {\tt mwdust} to the \Gaia{} bands using coefficients from \citet{Wang19} to compare the absolute magnitudes to the evolutionary tracks. The distance and extinction are kept fixed at the values reported in Table \ref{tab:summary_table}. We run the MCMC chains for 5000 iterations and use a burn-in of 500 iterations. The CMD position of the two targets suggest $M_1=\,$\PrimaryMass{J1208}$\ M_\odot$ for J1208 and $M_1=\,$\PrimaryMass{J1721}$\ M_\odot$ for J1721.


We also estimated the photometric primary masses using the constraints on $T_{\rm{eff}}$ and $R_1$ from the SED models. The resulting posteriors are much narrower but within the uncertainties of the previous model, with masses of $M_1=0.73\pm0.02\ M_\odot$ for J1208 and $M_1=0.793\pm0.005\ M_\odot$ for J1721, consistent with the CMD-only results given their uncertainties. Since the SED fits assumed a fixed extinction and distance, we report the broader posterior estimates without the SED priors in Table \ref{tab:summary_table}. Both of these primary mass estimates assume that the photometric primaries in both systems have evolved without any mass transfer history, which may not be reasonable given the nature of both of these systems as post-CE binaries. 

\begin{figure}
    \centering
    \includegraphics[width=\linewidth]{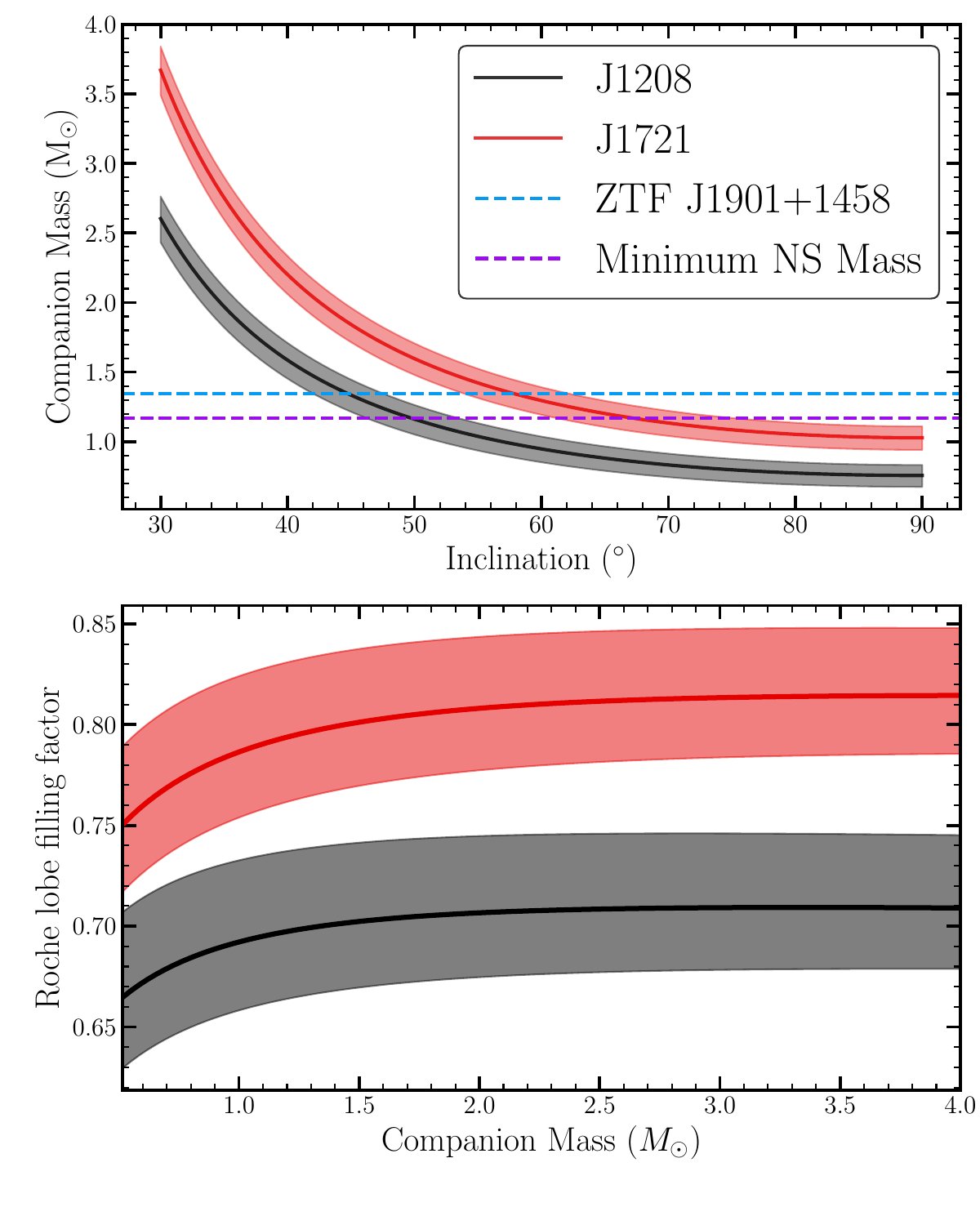}
    \caption{Top: Companion mass as a function of orbital inclination for J1208 and J1721 from the radial velocity constraints and the estimated primary mass. The shaded region shows a $0.15\ M_\odot$ uncertainty. The horizontal blue line shows the mass of the most massive known WD, ZTF~J1901$+$1458 \citep[$M=1.35\ M_\odot$,][]{Caiazzo21}. The purple line shows the theoretical minimum neutron star mass \citep[$M=1.17\ M_\odot$,][]{Suwa18}. Bottom: Roche lobe filing factors (Equation \ref{eqn:Eggleton}) for the range of $M_2$ in the top panel. Both targets have $f<1$ for this range of companion masses.}
    \label{fig:incl_m2}
\end{figure}

Figure \ref{fig:incl_m2} shows the constraints on the companion mass from the radial velocity observations assuming these K-dwarf mass values. Both targets are consistent with massive WDs for a broad range of inclinations. However, for inclinations less than $\approx 43^{\circ}$ and $\approx 55^{\circ}$, for J1208 and J1721, respectively, the companion masses would exceed the Chandrasekhar limit and they would have to be neutron stars. If we take the limiting case and assume that the orbital inclinations are distributed uniformly in $\cos i$, this corresponds to a probability of $73\%$ and $57\%$ that J1208 and J1721 have $M_2 < 1.4\ M_\odot$, respectively.

We can also determine the range of Roche-lobe filling factors $f=R_1/R_{\rm{RL}}$ for different values of the companion mass $M_2$. We estimate the Roche lobe radius $R_{\rm{RL}}$ as \citep{Eggleton83}
\begin{equation} \label{eqn:Eggleton}
    \frac{R_{\rm{RL}}}{a} = \frac{0.49 q^{-2/3}}{0.6q^{-2/3}+\ln(1+q^{1/3})},
\end{equation}
\noindent where $a$ is the semimajor axis and $q=M_2/M_1$, where $M_1$ is the K-dwarf mass. The bottom panel of Figure \ref{fig:incl_m2} shows the filling factors for the two targets. Neither are close to filling their Roche lobes ($f<1$), suggesting there is no ongoing mass transfer.

\section{Spotted Ellipsoidal Light Curve Fits} \label{sec:lcs}

We use the \TESS{} light curves to determine the orbital periods of the binary and fit spectroscopic orbits in Section \S\ref{sec:characterization}. Ellipsoidal modulations can also be used to constrain the mass ratio and inclination of the binary when fit simultaneously with the radial velocities \citep{Morris93}. 

Unfortunately, the light curves of J1208 and J1721 include additional variability due to spots. These vary between \TESS{} sectors and introduce asymmetric maxima in the light curves, as well as departures from symmetry around conjunction. Light curve modelling tools such as {\tt ELC} \citep{Orosz00} and \PHOEBE{} \citep{Prsa05, Conroy20} can include star spots in their light curve models, but this has only been done for a handful of targets \citep[e.g.,][]{Strader19, Lin23}. These models typically do not include prescriptions for time-dependent spot evolution. 

Here we attempt to model the \TESS{} light curves of J1208 and J1721 as spotted ellipsoidal variables with \PHOEBE{} to determine the mass ratio and binary inclination. However, there are degeneracies in the solutions of rotational variable light curves \citep{Luger21}, especially when using a single band light curve. We fit each \TESS{} sector independently, testing one-spot and a two-spot models. We simultaneously fit the light curve data and the radial velocities. 

For all \PHOEBE{} models, we treat the secondary as a dark companion by fixing it to be small, $R_2=3\times 10^{-6}\ R_\odot$, and cold, $T_{\rm{eff,2}}=300$~K. We do not include the effects of irradiation or reflection, as in \citet{Jayasinghe22}. We also fix the eccentricity $e=0$ based on the RV fit. Each spot on the K-dwarf has an independent latitude, $\theta_s$, longitude, $\phi_s$, angular size, $R_s$, and temperature, $T_s$, parameterized as a relative temperature, $T_s/T_{\rm{eff}}$. The latitude is defined such that $\theta_s=0^{\circ}$ occurs at the pole corresponding to the spin axis, and $\phi_s=0^{\circ}$ corresponds to the direction facing the companion. 

We start by using the differential evolution optimizer in \PHOEBE{} to identify an initial state for the MCMC sampling. We set Gaussian priors on the effective temperature ($\sigma=100$~K), and primary radius ($\sigma=0.1~R_\odot$) based on the SED fits, and a uniform prior on the primary mass ($[0.5~M_\odot,~0.95~M_\odot]$). We run each \PHOEBE{} model for 10000 iterations with 16 walkers for the one-spot models and 20 walkers for the two-spot models. We inspected the walker distributions to select suitable burn-in periods, typically 2000 -- 5000 iterations. 

{\renewcommand{\arraystretch}{1.2}
\begin{table*}
    \centering
    \caption{{\tt PHOEBE} posteriors for J1208. We fit each sector independently and use models with one and two spots.}
    \begin{tabular}{lllll}
\toprule
{} &                            S22, 1 spot &                            S22, 2 spot &                         S49, 1 spot &                         S49, 2 spot \\
\midrule
$\rm{Period}\ (d)$         &  $0.4632676^{+0.0000005}_{-0.0000005}$ &  $0.4632672^{+0.0000008}_{-0.0000006}$ &  $0.463250^{+0.000002}_{-0.000002}$ &  $0.463322^{+0.000003}_{-0.000003}$ \\
$T_0$                      &      $2459984.8070^{+0.001}_{-0.0009}$ &        $2459984.807^{+0.001}_{-0.001}$ &   $2459984.8073^{+0.001}_{-0.0010}$ &     $2459984.802^{+0.001}_{-0.001}$ \\
$q$                        &                    $1.5^{+0.1}_{-0.1}$ &                 $1.26^{+0.07}_{-0.06}$ &               $1.27^{+0.06}_{-0.1}$ &              $1.44^{+0.06}_{-0.06}$ \\
$i\ (^{\circ})$            &                         $62^{+7}_{-4}$ &                         $73^{+3}_{-3}$ &                      $90^{+5}_{-5}$ &                      $56^{+1}_{-1}$ \\
$R_1\ (R_\odot)$           &                 $0.77^{+0.07}_{-0.05}$ &                 $0.74^{+0.03}_{-0.03}$ &              $0.66^{+0.05}_{-0.02}$ &              $0.73^{+0.02}_{-0.02}$ \\
$T_{\rm{eff},1}\ (\rm{K})$ &                     $4710^{+80}_{-90}$ &                   $4700^{+200}_{-100}$ &                  $4670^{+70}_{-70}$ &                  $4740^{+50}_{-50}$ \\
$\gamma\ (\rm{km/s})$      &                         $-7^{+1}_{-3}$ &                   $-6.1^{+0.7}_{-0.6}$ &                      $-7^{+1}_{-1}$ &                $-5.0^{+0.1}_{-0.1}$ \\
$M_1\ (M_\odot)$           &                  $0.60^{+0.1}_{-0.08}$ &                 $0.68^{+0.08}_{-0.07}$ &               $0.55^{+0.1}_{-0.04}$ &              $0.78^{+0.04}_{-0.05}$ \\
$M_2\ (M_\odot)$           &                    $0.9^{+0.1}_{-0.1}$ &                 $0.86^{+0.07}_{-0.07}$ &              $0.71^{+0.05}_{-0.03}$ &              $1.13^{+0.03}_{-0.05}$ \\
$\theta_{s,1}\ (^{\circ})$ &                         $40^{+6}_{-4}$ &                        $133^{+2}_{-2}$ &                     $120^{+5}_{-5}$ &                $75.8^{+0.8}_{-0.7}$ \\
$\phi_{s,1}\ (^{\circ})$   &                         $93^{+2}_{-1}$ &                         $63^{+1}_{-2}$ &                      $36^{+2}_{-1}$ &               $315.9^{+0.3}_{-0.3}$ \\
$R_{s,1}\ (^{\circ})$      &                         $18^{+2}_{-1}$ &                        $-2^{+7}_{-10}$ &                      $33^{+4}_{-2}$ &                $27.9^{+0.4}_{-0.3}$ \\
$T_{s,1}/T_{\rm{eff}}$     &                 $0.81^{+0.04}_{-0.04}$ &                  $0.80^{+0.09}_{-0.3}$ &              $0.93^{+0.01}_{-0.01}$ &              $0.83^{+0.01}_{-0.01}$ \\
$\theta_{s,2}\ (^{\circ})$ &                                        &                      $9.5^{+2}_{-0.8}$ &                                     &                $21.2^{+0.3}_{-0.2}$ \\
$\phi_{s,2}\ (^{\circ})$   &                                        &                         $98^{+2}_{-2}$ &                                     &               $107.9^{+0.3}_{-0.4}$ \\
$R_{s,2}\ (^{\circ})$      &                                        &                         $42^{+2}_{-1}$ &                                     &                $54.9^{+0.6}_{-0.3}$ \\
$T_{s,2}/T_{\rm{eff}}$     &                                        &                 $0.85^{+0.06}_{-0.05}$ &                                     &           $0.942^{+0.002}_{-0.002}$ \\
\bottomrule
\end{tabular}

    \label{tab:j1208_phoebe}
\end{table*}}

{\renewcommand{\arraystretch}{1.2}
\begin{table*}
    \centering
    \caption{Same as Table \ref{tab:j1208_phoebe}, but for J1721.}
    \begin{tabular}{lllllll}
\toprule
{} &                    S12, 1 spot &                   S12, 2 spot &                         S13, 1 spot &                   S13, 2 spot &                        S39, 1 spot &                       S39, 2 spot \\
\midrule
$\rm{Period}\ (d)$         &   $0.4468^{+0.0004}_{-0.0003}$ &  $0.4472^{+0.0001}_{-0.0003}$ &  $0.4472498^{+0.0000005}_{-0.0003}$ &  $0.4471^{+0.0003}_{-0.0004}$ &   $0.446951^{+0.0003}_{-0.000001}$ &  $0.446637^{+0.0003}_{-0.000001}$ \\
$T_0$                      &  $2459379.536^{+0.002}_{-0.1}$ &   $2459379.20^{+0.2}_{-0.03}$ &    $2459379.3922^{+0.0006}_{-0.03}$ &   $2459379.18^{+0.2}_{-0.03}$ &  $2459379.3711^{+0.003}_{-0.0006}$ &   $2459379.364^{+0.003}_{-0.001}$ \\
$q$                        &          $1.62^{+0.5}_{-0.05}$ &         $1.95^{+0.2}_{-0.06}$ &                 $1.9^{+0.1}_{-0.1}$ &        $2.24^{+0.06}_{-0.08}$ &             $1.92^{+0.09}_{-0.07}$ &            $1.36^{+0.05}_{-0.04}$ \\
$i\ (^{\circ})$            &                 $56^{+3}_{-4}$ &                $44^{+2}_{-3}$ &                      $55^{+5}_{-5}$ &                $45^{+3}_{-2}$ &                     $49^{+1}_{-1}$ &                    $67^{+1}_{-1}$ \\
$R_1\ (R_\odot)$           &         $0.85^{+0.02}_{-0.05}$ &        $0.91^{+0.02}_{-0.03}$ &              $0.80^{+0.07}_{-0.03}$ &        $0.88^{+0.01}_{-0.01}$ &             $0.90^{+0.02}_{-0.02}$ &         $0.888^{+0.009}_{-0.008}$ \\
$T_{\rm{eff},1}\ (\rm{K})$ &             $5080^{+90}_{-50}$ &            $5110^{+90}_{-30}$ &                $5200^{+100}_{-100}$ &           $5080^{+200}_{-60}$ &                 $5130^{+50}_{-70}$ &                $5040^{+50}_{-50}$ \\
$\gamma\ (\rm{km/s})$      &            $-38.7^{+2}_{-0.7}$ &         $-39.6^{+0.2}_{-0.2}$ &               $-34.1^{+0.7}_{-0.8}$ &           $-34.0^{+0.4}_{-1}$ &                $-47.4^{+1}_{-0.9}$ &             $-40.3^{+0.2}_{-0.2}$ \\
$M_1\ (M_\odot)$           &          $0.84^{+0.01}_{-0.1}$ &        $0.88^{+0.06}_{-0.04}$ &               $0.73^{+0.1}_{-0.06}$ &        $0.86^{+0.02}_{-0.03}$ &             $0.70^{+0.03}_{-0.04}$ &          $0.838^{+0.007}_{-0.01}$ \\
$M_2\ (M_\odot)$           &          $1.43^{+0.1}_{-0.10}$ &        $1.76^{+0.09}_{-0.06}$ &                 $1.3^{+0.3}_{-0.1}$ &        $1.89^{+0.08}_{-0.06}$ &             $1.36^{+0.06}_{-0.07}$ &            $1.15^{+0.04}_{-0.05}$ \\
$\theta_{s,1}\ (^{\circ})$ &             $52.2^{+2}_{-0.4}$ &          $45.7^{+0.6}_{-0.8}$ &                  $16.3^{+2}_{-0.9}$ &               $120^{+1}_{-1}$ &               $71.8^{+0.8}_{-0.8}$ &                   $228^{+2}_{-3}$ \\
$\phi_{s,1}\ (^{\circ})$   &             $89.0^{+2}_{-0.9}$ &                $89^{+1}_{-2}$ &                  $89.9^{+0.7}_{-2}$ &          $55.2^{+0.7}_{-0.8}$ &              $210.5^{+0.8}_{-0.6}$ &                   $100^{+5}_{-9}$ \\
$R_{s,1}\ (^{\circ})$      &             $22.9^{+0.8}_{-1}$ &            $28.9^{+0.3}_{-1}$ &                      $29^{+2}_{-1}$ &          $10.9^{+0.8}_{-0.8}$ &               $16.1^{+0.6}_{-1.0}$ &                    $19^{+2}_{-2}$ \\
$T_{s,1}/T_{\rm{eff}}$     &         $0.87^{+0.01}_{-0.01}$ &      $0.906^{+0.005}_{-0.01}$ &              $0.86^{+0.03}_{-0.01}$ &        $0.81^{+0.05}_{-0.06}$ &          $0.780^{+0.007}_{-0.008}$ &            $0.95^{+0.02}_{-0.01}$ \\
$\theta_{s,2}\ (^{\circ})$ &                                &           $119.3^{+0.9}_{-2}$ &                                     &           $9.8^{+0.7}_{-0.4}$ &                                    &                    $38^{+3}_{-3}$ \\
$\phi_{s,2}\ (^{\circ})$   &                                &           $214.0^{+0.6}_{-2}$ &                                     &          $91.7^{+0.9}_{-0.8}$ &                                    &                   $198^{+2}_{-2}$ \\
$R_{s,2}\ (^{\circ})$      &                                &            $15.4^{+0.6}_{-2}$ &                                     &          $38.0^{+0.3}_{-0.4}$ &                                    &              $19.1^{+0.6}_{-0.9}$ \\
$T_{s,2}/T_{\rm{eff}}$     &                                &         $0.76^{+0.1}_{-0.04}$ &                                     &      $0.876^{+0.006}_{-0.01}$ &                                    &            $0.81^{+0.03}_{-0.02}$ \\
\bottomrule
\end{tabular}

    \label{tab:j1721_phoebe}
\end{table*}}

\begin{figure*}
    \centering
    \includegraphics[width=\linewidth]{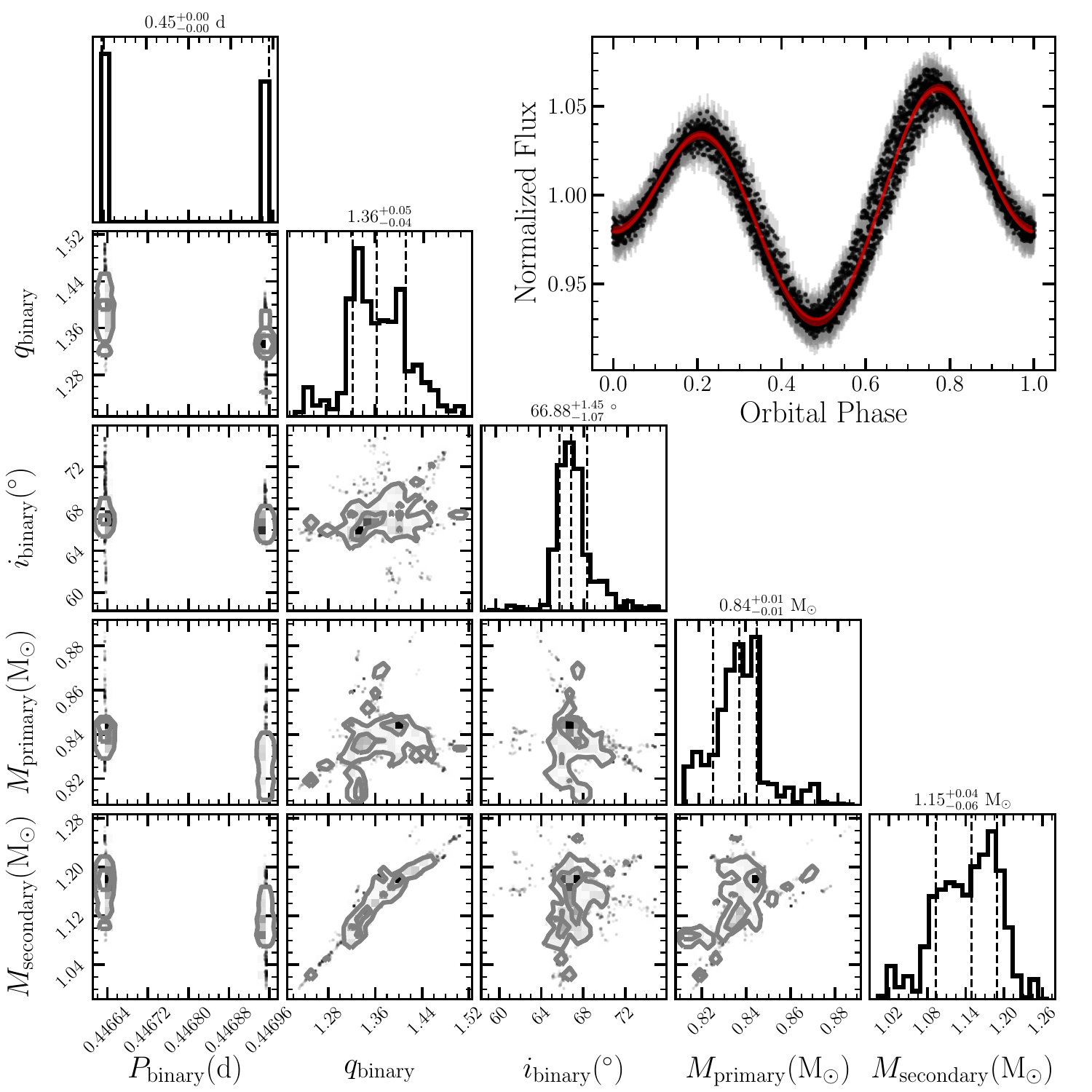}
    \caption{MCMC posteriors for the sector 39 two-spot model of J1721. The light curve fit is shown in the upper right. This model predicts a companion mass consistent with a massive white dwarf.}
    \label{fig:j1721_corner_example}
\end{figure*}

\begin{figure}
    \centering
    \includegraphics[width=\linewidth]{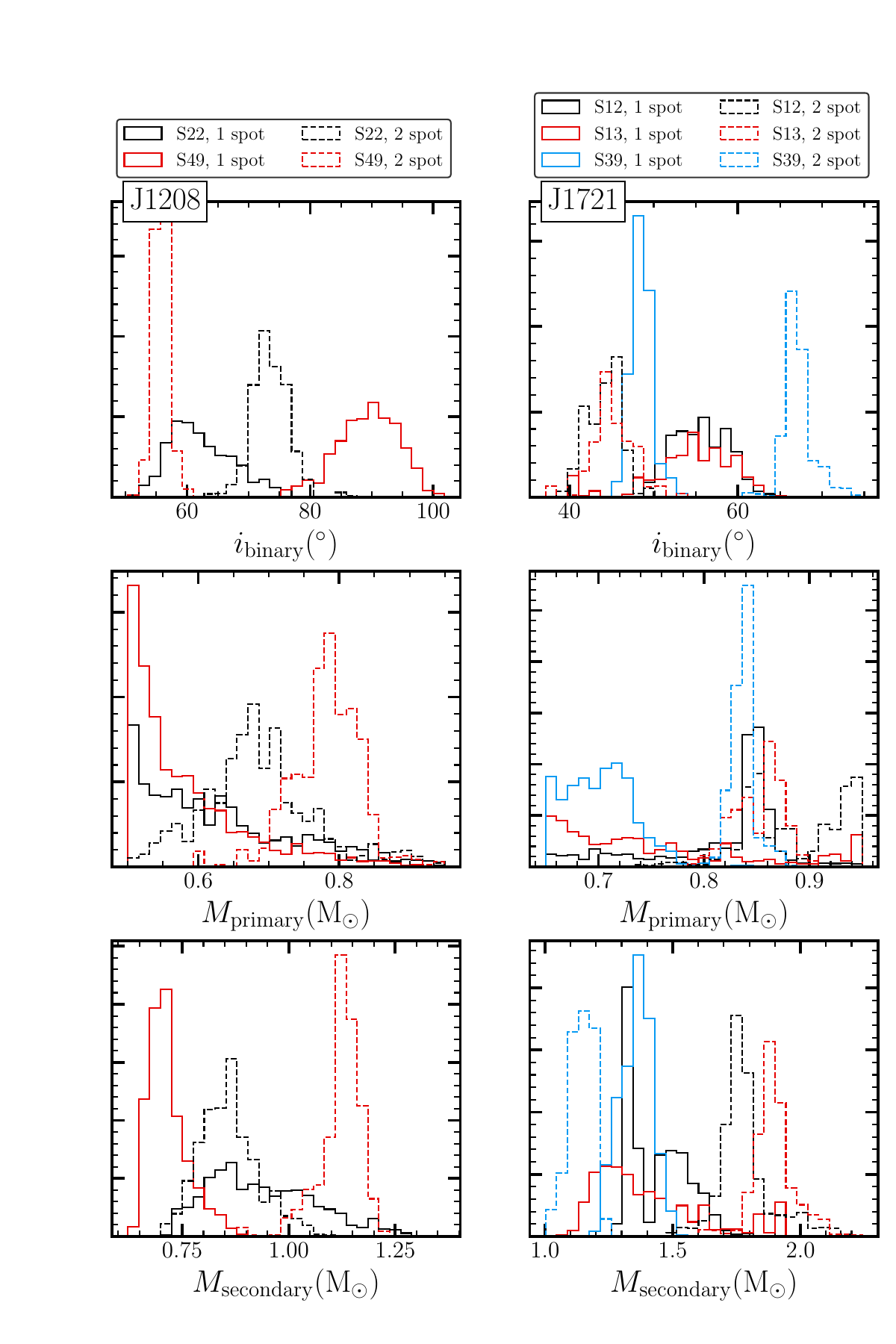}
    \caption{MCMC posteriors for J1208 (left) and J1721 (right). We find that the different sectors and number of spots produces variable predictions of the orbital inclination and secondary mass.}
    \label{fig:mcmc_histograms}
\end{figure}

Table \ref{tab:j1208_phoebe} reports the MCMC posteriors for the two sectors of J1208 and Table \ref{tab:j1721_phoebe} reports the results for J1721. Figure \ref{fig:j1721_corner_example} shows an example corner plot and light curve fit for J1721 with a one-spot model for sector 12. Figure \ref{fig:mcmc_histograms} compares the inclination, primary mass, and secondary mass posteriors for the different model fits. For both targets, we find that the models do not produce a consistent prediction for the companion mass between the \TESS{} sectors and for different spot models.

Three of the four \PHOEBE{} models of J1208 predict K-dwarf masses below what is expected based on the SED and CMD. While this could suggest a history of mass-transfer, it seems unlikely that $\sim0.2\ M_\odot$ of material was transferred to the white dwarf companion. The sector 22 two-spot model also finds a spot size consistent with zero for one of the spots, which may suggest a 1-spot model is preferable for that sector. However, we note that the position and size of the spot differs between the two models (Table \ref{tab:j1208_phoebe}). 

We can also compare the predicted spot temperatures to expectations based on analytic models from \citet{Berdyugina05}, where the spot temperature $T_s$ is related to the effective temperature by

\begin{equation} \label{eqn:spot_temp}
    T_s = -895\left(\frac{T_{\rm{eff}}}{5000\ \rm{K}}\right)^{2} + 3755 \left(\frac{T_{\rm{eff}}}{5000\ \rm{K}}\right) + 808\ \rm{K}
\end{equation}

\noindent Based on the effective temperature from the SED (\SEDTeff{J1208}~K), the spot temperature is predicted to be \AnalyticSpotTeff{J1208}~K, which corresponds to a ratio $T_s/T_{\rm{eff}}=\AnalyticSpotTeffRatio{J1208}$. This value is lower than the MCMC results for all models. This could indicate that this target has an atypically low contrast between the spot and the photosphere, or that a more complex spot model is necessary for J1208.

\begin{figure}
    \centering
    \includegraphics[width=\linewidth]{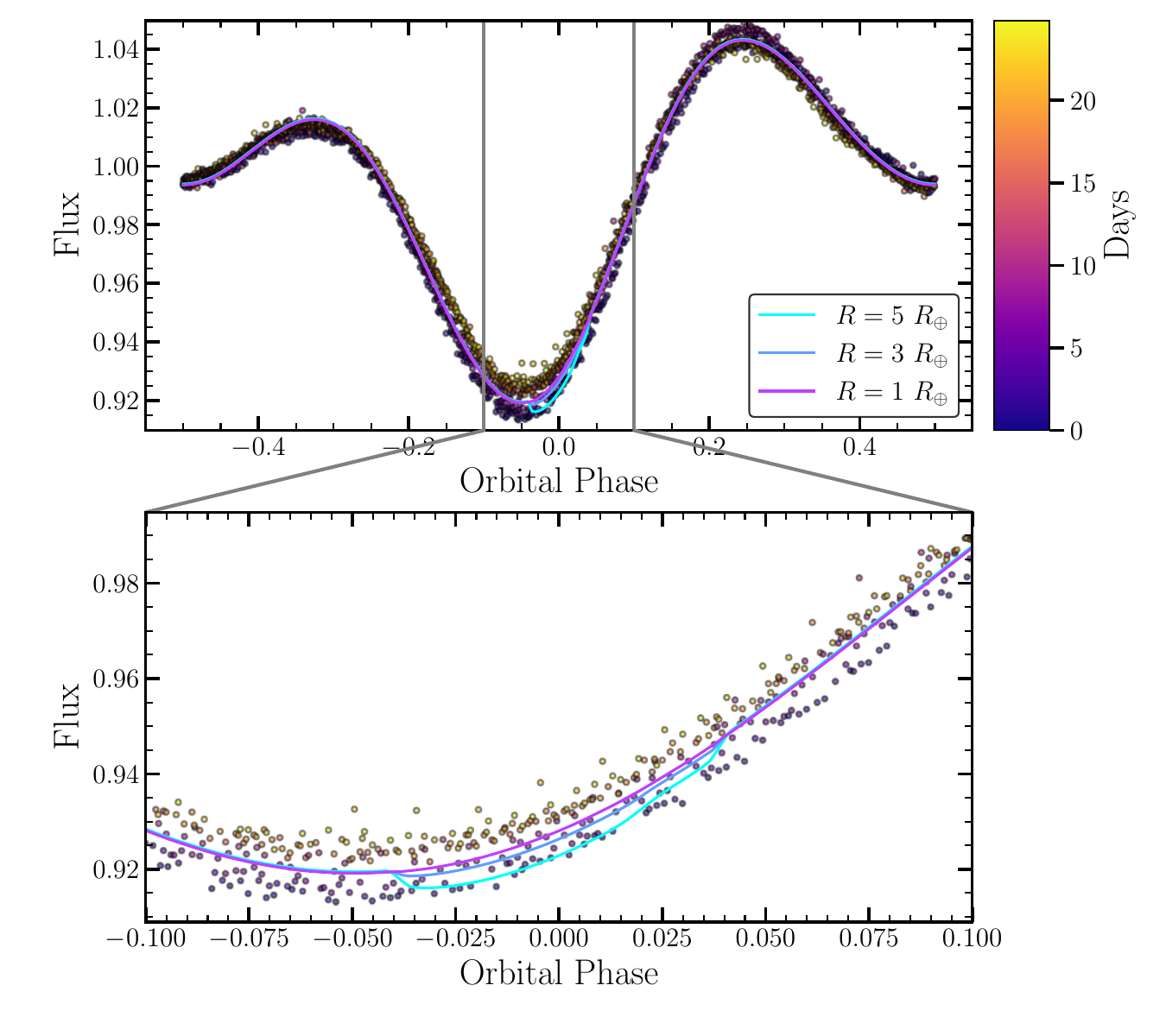}
    \caption{The \PHOEBE{} sector 49 one-spot model shown for different WD radii. The sector 49 one-spot model (Table \ref{tab:j1208_phoebe}) prefers an edge-on inclination and a WD mass that corresponds to a radius $R_2\approx 0.95\ R_\oplus$ based on WD scaling relation. However, we are unable to confirm or reject this model on the basis of eclipse detection, in part due to the scatter in the model from the spot modulation throughout the \TESS{} sector. To center the eclipse feature at phase $\phi=0$, orbital phase is defined such that RV maxima occurs at $\phi=-0.25$ in this figure.}
    \label{fig:j1208_eclipse_constraint}
\end{figure}

The sector 49 models both find larger spots, with similar temperatures and sizes. The sector 49 two-spot model predicts a higher primary mass ($M_1=0.78^{+0.04}_{-0.05}\ M_\odot$) and a higher mass ratio, $q=1.44^{+0.06}_{-0.06}$. The result is a much larger white dwarf mass, $M_2=1.13^{+0.03}_{-0.05}$.  While it is tempting to prefer this model because of the better agreement with the SED mass and radius, it is clear that there are numerous degeneracies in the light curve solutions that limit our ability to characterize the WD.

The sector 49 one-spot model also predicts an edge-on inclination, $i=90\pm5^\circ$. We might expect to be able to rule this model out based on the lack of an eclipse. Using the non-relativistic WD scaling relation $R\propto M^{-1/3}$, a WD mass of $M_2=0.71\ M_\odot$ should have a radius $R_2\approx 0.95\ R_\oplus$. Figure \ref{fig:j1208_eclipse_constraint} shows the best fit \PHOEBE{} model after changing $R_2$ to be consistent with a WD of a few Earth radii. The scatter around the model primarily comes from the evolution of the spot across the \TESS{} sector. Although we can plausibly rule out a companion of radius $R_2=5\ R_\oplus$, we would not expect to detect eclipses of companions $R_2\sim 1\ R_\oplus$ and therefore cannot place an upper limit on the inclination.

In all \PHOEBE{} models of J1721, we find multimodal posterior distributions for the orbital period, $P$. The difference between the periods is small, typically 15 -- 45 minutes. This could come from slightly asynchronous rotation, where the orbital period does not equal the rotation period, or latitudinal differential rotation in the K-dwarf. Similar differences were found between the orbital period and photometric period of the G+WD binary CPD-65~264 \citep{Hernandez22b}.

For J1721, the secondary mass posteriors span from $1$--$1.9\ M_\odot$ depending on the sector and number of spots. For the sector 13 and 39 single spot models, the primary mass posterior is lower than what is expected from the SED. The two-spot model for sector 12 predicts a primary mass at the upper range of our mass prior, $\sim 0.95\ M_\odot$. The two-spot models for sector 13 and 39 predict primary masses consistent with our expectations from the CMD, but yield discrepant inclinations $45^{+3^\circ}_{-2^\circ}$ and $67^{+1^\circ}_{-1^\circ}$, respectively. The sector 13 model then predicts a companion with $M_2=1.89^{+0.08}_{-0.06}\ M_\odot$, consistent with a neutron star companion, while sector 39 model predicts $M_2=1.15^{+0.04}_{-0.05}\ M_\odot$, consistent with a massive white dwarf.

In comparing the spot parameters for these two models, the sector 13 two-spot model predicts two spots of different angular sizes with similar temperatures. The sector 39 two-spot model predicts two spots of the same size, but with different temperatures. Since the spots evolve, it is not surprising that the positions and sizes of the spots are different, especially since the two sectors are separated by almost two years. As with J1208, the posterior spot temperature ratio is higher than what is predicted from the analytic model, $T_s/T_{\rm{eff}}=\AnalyticSpotTeffRatio{J1721}$. 

Unlike J1208, where the two \TESS{} sectors are separated by ${\sim2}$~years, J1721 was observed in two consecutive sectors (12 and 13). The light curves of these two sectors are similar, with the sector 13 observations showing a higher second maxima. In the one-spot models of these two sectors, the effective temperature of the spots are similar, but the size increases going from sector 12 to sector 13. Models of both sectors prefer an inclination $\sim 55^{\circ}$, and the spots in the two models appear at similar longitudes. The sector 13 spot prefers a lower latitude. 

In comparing the two-spot models of sector 12 and 13, we see that spot one of the sector 12 model shares a similar longitude and temperature with spot two of the sector 13 model, though it appears at a different latitude and gets larger by $\sim10^{\circ}$. Alternatively, the second spot of the sector 12 model appears at a similar latitude to the first spot of the sector 13 model, but with different longitude and a decreasing size $\sim 4.5^{\circ}$. An important caveat with the two-spot models of sectors 12 and 13 is that one of the spots (spot two of sector 12 and spot one of sector 13) appears at a latitude $\sim 120^{\circ}$, which is not visible to the observer in the plane of the sky given the orbital inclination. These spots therefore contribute negligibly to the light curve. This may indicate a preference for the one-spot model. While these models may offer insight into the magnetic fields of chromospherically active close binaries, there is no reason to prefer a model with two spots over a model with more spots. 

In summary, the light curves of both targets show ellipsoidal modulations with additional variability due to spots. Since the spots evolve over time, the \TESS{} light curve morphology changes between sectors. We attempt to use \PHOEBE{} to model the light curves using one and two-spot models, but degeneracies with spot parameters limit our ability to characterize the systems. In light of these results, we also perform an injection-recovery test for a synthetic spotted binary in Appendix \ref{sec:injection_recovery}.

\section{Discussion} \label{sec:discussion}

The \TESS{} light curves, radial velocity observations, and SEDs suggest that J1208 and J1721 are K-dwarf binaries with massive white dwarf companions. Whereas the majority of WD+MS binaries have M-dwarf stars, J1208 and J1721 have K-dwarf photometric primaries, of which there are relatively few systems known \citep{Wonnacott93, Hernandez22b, Zheng22a}. Both systems have large RV amplitudes and have magnitudes and colors consistent with a single-star isochrone (Figure \ref{fig:rvamp_cmd}). The \TESS{} light curves show periodic variability at $\sim 0.45$~days. The light curve morphology changes dramatically between \TESS{} sectors (Figure \ref{fig:tess_lcs_folded}) due to the combination of ellipsoidal variability and star spots. 

We obtain radial velocity observations of these targets using APF, PEPSI, and CHIRON (Tables \ref{tab:j1208_rvs} and \ref{tab:j1721_rvs}). The spectroscopic orbits (Figure \ref{fig:rv_orbits}) imply mass functions $f(M)=\BinaryMassFunction{J1208}\ M_\odot$ and $f(M)=\BinaryMassFunction{J1721}\ M_\odot$ for J1208 and J1721, respectively. We then use SEDs to estimate the primary mass of each target (Figure \ref{fig:seds}), which can be used to place broad constraints on the companion mass (Figure \ref{fig:incl_m2}), suggesting that these are massive WDs.

There are also some key differences between the two targets. J1208 has a more significant FUV excess and an X-ray detection. The X-ray to bolometric luminosity ratio is consistent with what is expected from chromospheric activity. This target also shows evidence of \Halpha{} emission (Figure \ref{fig:j1208_halpha}) and long-term photometric variability (Figure \ref{fig:archival_unfolded}), neither of which are observed for J1721. It could be the case that J1208 is more chromospherically active, or that J1721 is in a low activity state. 

Our ability to characterize the unseen companion using radial velocities is limited by the unknown orbital inclination. We attempt to model the \TESS{} light curves using \PHOEBE{} to estimate the inclination, treating each sector independently and using one and two-spot models. We find that the models do not predict a consistent orbital inclination and secondary mass between sectors, presumably due to degeneracies in modeling the light curves of spotted stars using a single photometric band. 

The J1208 \PHOEBE{} models generally prefer a low-mass photometric primary that is not consistent with the SED $T_{\rm{eff}}$ or radius. This could point to an accretion disk as the source of the \Halpha{} emission and X-ray variability, but such models are beyond the scope of this work. Models of the binary evolution history could also be used to place constraints on the amount of mass that could be transferred from the K-dwarf. The J1721 \PHOEBE{} models predict a range of companion masses, including some $>1.4\ M_\odot$ consistent with a neutron star companion. However, like J1208, the estimates of the primary mass do not match the expectations from the SED, making interpretation of these models challenging. 

Simultaneous multi-band light curves could be used to break some of the degeneracies in the light curves of rotational variability \citep{Luger21}. We are ultimately less interested in determining the spot parameters than in determining the orbital inclination and mass ratio, so Gaussian processes may provide a pathway to handling spot evolution and fitting multiple \TESS{} sectors simultaneously \citep[e.g.,][]{Luger21_II}. Since J1721 was observed in two consecutive sectors, this target may be a good test-case for time-dependent models. Ultraviolet spectroscopy seems to be the most promising approach to better constraint the nature of the companions, particularly to discriminate between neutron stars and white dwarfs. For example, \citet{Hernandez22b} characterized the WD companion ($M=0.86\pm0.06\ M_\odot$) to a G-dwarf using HST UV spectroscopy. Similar observations of J1208 and J1721 could provide additional constraints on the compact object companions.

We might also expect the kinematics of WD and neutron star binaries to differ. Neutron stars are expected to experience natal kicks following the supernova that can affect their subsequent motion through the Galaxy. X-ray binaries with neutrons stars have been found to have Galactic kinematics significantly different from ``normal'' stars \citep[e.g.,][]{Hernandez2005}. Natal kicks are expected to be $\approx 50\%$ larger for neutron stars than black holes \citep{Atri19, Odoherty23}, and we may expect J1208 and J1721 to have atypical Galactic orbits if they host a NS companion. We use the \Gaia{} DR3 parallax, proper motion, and the center-of-mass velocity from the RVs to estimate the trajectory of J1208 and J1721 in the Galaxy. We use {\tt galpy} \citep{Bovy15} and the {\tt MWPotential2014} potential to integrate the orbits from 500 Myr ago to 500 Myr in the future. Both orbits are consistent with the thin disk, staying $\lesssim 200$~pc from the Galactic midplane \citep{Du06}.

In recent years there have been a number of candidate non-interacting compact objects identified based on the photometric and RV variability of late-type stars. \citet{Zheng22b} reported LAMOST J235456.76+335625.7 (J2354) as a nearby neutron star candidate with mass $M_2 > 1.26\pm0.03\ M_\odot$. Like J1208 and J2354, this binary has an orbital period $P=0.47991$~days and a light curve dominated by spotted ellipsoidal variability in \TESS{} observations. J2354 also has a significant GALEX NUV excess and an \Halpha{} emission line that traces the motion of the K-dwarf. While \citet{Zheng22b} propose that the companion to J2354 is a neutron star, the system also may be a massive white dwarf (Tucker et al., in preparation).

\citet{Lin23} also reported the detection of a similar system 2MASS J15274848+3536572 (J1527) using LAMOST RVs. The orbital period is shorter, with $P=0.256$~days and they find that the K9-M0 primary has a mass $M_1=0.62\pm0.01\ M_\odot$. The light curve is again similar to J1208, J1721, and J2354, and appears to show similar modulations over time. The mass function of J1527 is $f(M)=0.131\pm0.002~M_\odot$. They also attempt a \PHOEBE{} fit to the $B$, $V$, $R$, and \TESS{} light curves, and report $M_2=0.98\pm0.03\ M_\odot$. Unlike J1208 and J2354, the \Halpha{} emission moves in antiphase with the photometric primary, which could indicate the presence of an accretion disk. If the companion is instead a neutron star, the lack of X-ray and $\gamma$-ray detection suggests it is part of the X-ray dim NS population. A neutron star with $M=0.98\ M_\odot$ challenges our understanding of core collapse supernovae, which are expected to yield remnants with minimum masses $\sim1.17\ M_\odot$ \citep{Suwa18}.

\begin{figure}
    \centering
    \includegraphics[width=\linewidth]{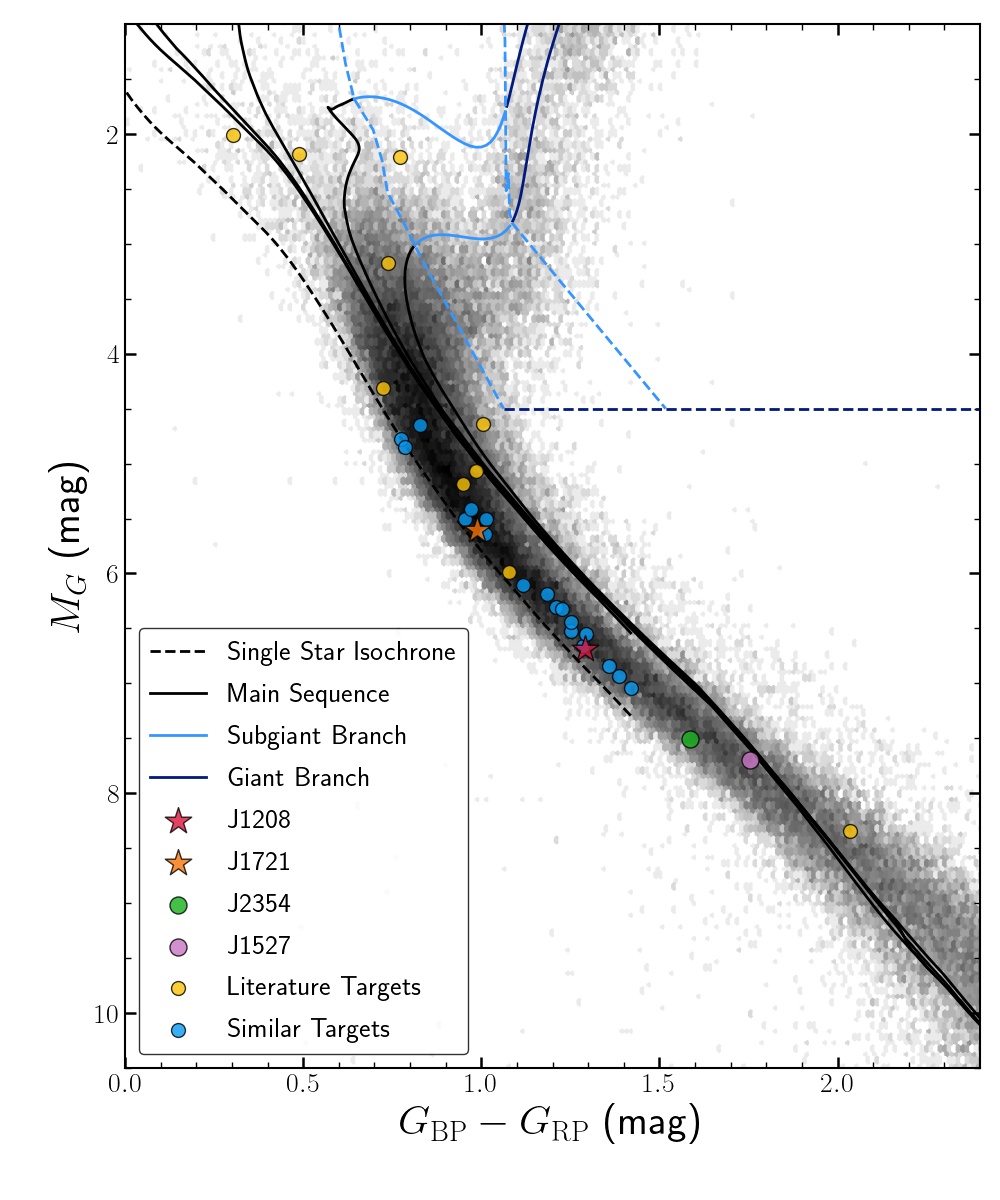}
    \caption{\Gaia{} color-magnitude diagram of our targets (J1208 and J1721) compared to similar targets identified in the literature (green). The solid black line shows a binary isochrone corresponding to an equal mass binary. We also identify a number of systems listed in Table \ref{tab:discussion_table} with similar light curves and high \Gaia{} $A_{\rm{RV}}$ shown in blue.}
    \label{fig:discussion_cmd}
\end{figure}

Figure \ref{fig:discussion_cmd} shows these systems on a \Gaia{} color-magnitude diagram. Similar targets have also been identified in \citet{Li22}, \citet{Qi23}, \citet{Fu22}, \citet{Zheng22a}, \citet{Hernandez22b}, and \citet{Yi22}, though some do not have evidence of rotational variability in their light curves. Some of these sources appear more luminous than expected for a single main sequence star. Spectra of these targets are needed to rule out a second luminous component.

If many of these objects turn out to be massive WDs, rather than luminous companions, this could have implications for the overall mass distribution of white dwarfs. J1208, J1721, J2354 \citep{Zheng22a}, and J1527 \citep{Lin23} are all within 250~pc using \Gaia{} distances \citep{BailerJones21}. The local density of K-dwarfs with massive WD companions is then at least $\rho_\odot = 3 N/4\pi R^3 \sim 6.1\times10^{-8}$~pc$^{-3}$. For a simple thin disk model normalized by this density with 
\begin{equation}
    \rho = \rho_\odot \exp\left( - (R-R_s)/R_d - |z|/h \right),
\end{equation}
\noindent where $R_d=3500$~pc is the disk scale length, $R_\odot$, $R_s=8500$~pc is the radius of the Sun from the Galactic center, $z$ is the distance from the Galactic midplane, and $h=150$~pc is the disk height, we can estimate that there are  $\gtrsim\num[group-separator={,}]{16000}$ such systems the Galaxy. This is far fewer than the number of NS expected in the Galaxy \citep[$\sim 10^8$--$10^{10}$,][]{Sartore10} and complicates the discrimination between NS and massive WD companions from a statistical perspective.

Finally, we also performed a simple search to identify other systems which may contain similar companions. We start by selecting stars with \Gaia{} $A_{RV}>50$~km/s, and $G<15$~mag that were flagged as photometric variables in \Gaia{} DR3. We also require that the \Gaia{} parallax error satisfies $\varpi/\sigma_\varpi > 5$ and that the $V$-band extinction is $A_V < 2.0$~mag. We then use the extinction-corrected \Gaia{} color-magnitude diagram to select stars that appear more consistent with a single main sequence star than a stellar binary. We do this by selecting targets fainter than a single star isochrone increased in luminosity by a factor of $1.5$ ($0.44$~mag). These selection criteria yield 826 targets in the absolute magnitude range $4.5 < M_G < 12$~mag. We then visually inspected their \TESS{} QLP light curves, when available, selecting systems with similar orbital periods and ELL/spotted ELL light curves. In total, we identified \nDiscTargets{} targets, which are shown in Figure \ref{fig:discussion_cmd} and listed in Table \ref{tab:discussion_table}. 

Table \ref{tab:discussion_table} also includes estimates of $f(M)$ using the photometric period and assuming $K=A_{\rm{RV}}/2$. While this may be useful as a way to prioritize targets, $A_{\rm{RV}}$ is overestimated for our two targets. We find $K\simeq 0.44 A_{\rm{RV}}$ and $K\simeq0.2 A_{\rm{RV}}$ for J1208 and J1721, respectively. Radial velocity observations of the targets in Table \ref{tab:discussion_table} are needed to identify luminous companions and constrain the binary mass function. It is likely that some of these objects are similar to the two targets described here or to the previous reported WD/NS candidates. 

J1208 and J1721 join a small population of FGK stars with massive WD companions. As compared to WD+M stars, there are relatively few WD+FGK binaries since the main sequence star outshines the WD at optical wavelengths. Spectroscopic orbits are promising tools to identify and characterize WD+K binaries. However, these searches are likely biased towards systems with large companion mass where the velocity semi-amplitude is large enough to rule out luminous companions. Upcoming spectroscopic missions such Milky Way Mapper \citep{Kollmeier17} and future \Gaia{} data releases are expected to expand the sample of WD+FGK binaries.

{\renewcommand{\arraystretch}{1.2}
\begin{table*}
    \centering
    \caption{Targets with similar CMD positions, light curves identified through \Gaia{} and \TESS{} in Section \S\ref{sec:discussion}. $M_G$ and $G_{\rm{BP}}-G_{\rm{RP}}$ report the extinction-corrected absolute magnitude and color, respectively. The mass function $f(M)$ is computed using the photometric period and treating $A_{\rm{RV}} = 2K$. For targets identified in previous work, we instead use the reported radial velocity semi-amplitude, $K$, when avaiable, to compute $f(M)$.}
    \sisetup{table-auto-round,
     group-digits=false}
    \setlength{\tabcolsep}{6pt}
    \begin{center}
        \begin{tabular}{l S[table-format=1.3] S[table-format=4.1] S[table-format=2.1] S[table-format=2.1] S[table-format=1.2] r r r r}
\toprule
      {GDR3 Source} &  {Period} &  {Distance} &     {$G$} &  {$M_G$} &  {$G_{\rm{BP}}-G_{\rm{RP}}$} &  {$A_{\rm{RV}}$} & {$K$} & {$f(M)$} & {Reference} \\
{} & {(d)} & {(pc)} & {(mag)} & {(mag)} & {(mag)} & {(km/s)} & {$(\rm{km/s})$} & {($M_\odot$)} & {}\\ 
\midrule
5379221348814190336 &  0.244329 &  211.462952 & 12.224591 & 5.502547 &                     1.014184 &           265.22 &     - &     0.06 &           - \\
 538490565647002752 &  0.254756 &  309.284088 & 12.865199 & 4.773744 &                     0.773373 &            96.14 &     - &  $<0.01$ &           - \\
 888901687304620416 &  0.271789 &  138.248932 & 12.638993 & 6.935684 &                     1.386844 &           165.28 &     - &     0.02 &           - \\
 386667182582268928 &  0.336922 &  160.036407 & 12.571529 & 6.550435 &                     1.293825 &           212.38 &     - &     0.04 &           - \\
 124151875142913152 &  0.375076 &  245.143860 & 12.907320 & 5.417380 &                     0.971541 &           218.40 &     - &     0.05 &           - \\
 421818053931122176 &  0.381958 &  189.259827 & 12.824433 & 6.439141 &                     1.252892 &            50.08 &     - &  $<0.01$ &           - \\
 806594102274873344 &  0.501581 &  143.932281 & 12.098151 & 6.307360 &                     1.208750 &           108.31 &     - &     0.01 &           - \\
3717122881328373632 &  0.516191 &   75.350815 & 10.570268 & 6.184828 &                     1.185923 &            68.08 &     - &  $<0.01$ &           - \\
2990513370894699392 &  0.576676 &  127.991974 & 11.857308 & 6.321394 &                     1.227760 &            68.91 &     - &  $<0.01$ &           - \\
  17904898318910080 &  0.588986 &  205.427795 & 12.616076 & 5.502595 &                     0.954732 &           108.53 &     - &     0.01 &           - \\
4831037389874593920 &  0.696708 &  172.845840 & 13.052012 & 6.845022 &                     1.358981 &           291.35 &     - &     0.22 &           - \\
1842647931849558272 &  0.745748 &  159.734238 & 12.678954 & 6.661964 &                     1.284475 &            87.85 &     - &     0.01 &           - \\
 517420310786820352 &  1.013753 &  265.975525 & 12.696167 & 4.852485 &                     0.786578 &           296.93 &     - &     0.34 &           - \\
  65626554825500672 &  1.684941 &  204.862473 & 12.596542 & 5.637171 &                     1.011639 &            50.10 &     - &  $<0.01$ &           - \\
5570313476125894528 &  1.739302 &  388.706482 & 12.696926 & 4.650250 &                     0.826697 &           259.00 &     - &     0.39 &           - \\
5702192817771804416 &  1.859799 &  169.964844 & 12.256246 & 6.104451 &                     1.118523 &           348.88 &     - &     1.02 &           - \\
2052252132725589888 &  2.543089 &  192.014145 & 12.969660 & 6.520600 &                     1.252131 &            75.75 &     - &     0.01 &           - \\
2809553337714753920 &  5.678659 &  133.043991 & 12.657634 & 7.037658 &                     1.420611 &            51.90 &     - &     0.01 &           - \\
\midrule
\textbf{Literature Targets:} & & & & & & & & \\ 
1375051479376039040 &      0.26 &  117.975426 & 13.054289 & 7.695331 &                     1.754247 &          358.05 & $171.09^{+1.0}_{-0.97}$ &     0.13 &        \citet{Lin23} \\
1542461401838152960 &      0.27 &  352.493530 & 12.799067 & 5.063311 &                     0.986489 &          726.78 &                       - &     1.34 &         \citet{Fu22} \\
 770431444010267392 &      0.27 &  314.473907 & 15.832041 & 8.344118 &                     2.035088 &               - &           $257.0\pm2.0$ &     0.47 &         \citet{Yi22} \\
5687390848640176000 &      0.45 &  363.837830 & 12.505296 & 4.637274 &                     1.005107 &          590.24 &                       - &     1.20 &         \citet{Fu22} \\
2874966759081257728 &      0.48 &  127.261543 & 13.025951 & 7.502465 &                     1.585217 &          458.46 &           $219.4\pm0.5$ &     0.53 &     \citet{Zheng22b} \\
1633051023841345280 &      0.60 &  229.506454 & 12.942164 & 5.990063 &                     1.078550 &               - &   $124.4^{+1.0}_{-1.1}$ &     0.12 &     \citet{Zheng22a} \\
3379200833078092672 &      0.69 & 2798.946040 & 14.544648 & 2.013422 &                     0.302947 &               - &         $133.22\pm0.47$ &     0.17 &         \citet{Qi23} \\
 608189290627289856 &      0.79 &  324.910400 & 12.767137 & 5.187158 &                     0.948391 &          347.97 & $128.4^{+0.84}_{-0.73}$ &     0.17 &         \citet{Li22} \\
3381426828726340992 &      1.23 & 1327.539920 & 13.892116 & 3.171073 &                     0.737637 &               - &           $99.1\pm0.64$ &     0.12 &         \citet{Qi23} \\
4672702561514172544 &      1.37 &  205.238098 & 10.964239 & 4.312936 &                     0.725459 &          202.09 &         $100.83\pm0.09$ &     0.15 & \citet{Hernandez22b} \\
3381441465975139968 &      2.91 & 2246.095210 & 13.958452 & 2.180150 &                     0.488671 &               - &           $67.18\pm0.4$ &     0.09 &         \citet{Qi23} \\
  66844160873323008 &      3.93 &  841.156189 & 12.240489 & 2.214047 &                     0.770942 &          133.35 &           $61.72\pm0.3$ &     0.10 &         \citet{Qi23} \\
\bottomrule
\end{tabular}

    \end{center}
    \label{tab:discussion_table}
\end{table*}
}

\section*{Acknowledgements}

We thank the \PHOEBE{} developers for their support in implementing various light curve models. We thank Las Cumbres Observatory and its staff for their continued support of ASAS-SN. ASAS-SN is funded in part by the Gordon and Betty Moore Foundation through grants GBMF5490 and GBMF10501 to the Ohio State University, and also funded in part by the Alfred P. Sloan Foundation grant G-2021-14192.

This work presents results from the European Space Agency space mission Gaia. Gaia data are being processed by the Gaia Data Processing and Analysis Consortium (DPAC). Funding for the DPAC is provided by national institutions, in particular the institutions participating in the Gaia MultiLateral Agreement.

This paper includes data collected with the \textit{TESS} mission, obtained from the MAST data archive at the Space Telescope Science Institute (STScI). Funding for the TESS mission is provided by the NASA Explorer Program. STScI is operated by the Association of Universities for Research in Astronomy, Inc., under NASA contract NAS 5-26555. CSK and DMR TESS research is supported by NASA grant 80NSSC22K0128.

Support for TJ and DVM was provided by NASA through the NASA Hubble Fellowship grants HF2-51509 and HF2-51464 awarded by the Space Telescope Science Institute, which is operated by the Association of Universities for Research in Astronomy, Inc., for NASA, under contract NAS5-26555. C.Y.L. acknowledges support from NASA FINESST grant No. 80NSSC21K2043 and a research grant from the H2H8 Foundation.

The LBT is an international collaboration among institutions in the United States, Italy, and Germany. LBT Corporation partners are: The University of Arizona on behalf of the Arizona Board of Regents; Istituto Nazionale di Astrofisica, Italy; LBT Beteili- gungsgesellschaft, Germany, representing the Max-Planck Society, The Leibniz Institute for Astrophysics Potsdam, and Heidelberg University; The Ohio State University, representing OSU, University of Notre Dame, University of Minnesota, and University of Virginia. PEPSI was made possible by funding through the State of Brandenburg (MWFK) and the German Federal Ministry of Education and Research (BMBF) through their Verbundforschung grants 05AL2BA1/3 and 05A08BAC.

\section*{Data Availability}

The \textit{Gaia} DR3 data and the ASAS-SN and \TESS{} light curves are all publicly available. 

The ASAS-SN photometric data underlying this article are available in the the ASAS-SN Photometry Database (https://asas-sn.osu.edu/photometry). The data underlying this article are available in the article and in its online supplementary material.



\bibliographystyle{mnras}
\bibliography{massivewds} 




\appendix

\section{An injection-recovery test} \label{sec:injection_recovery}

\begin{figure*}
    \centering
    \includegraphics[width=\linewidth]{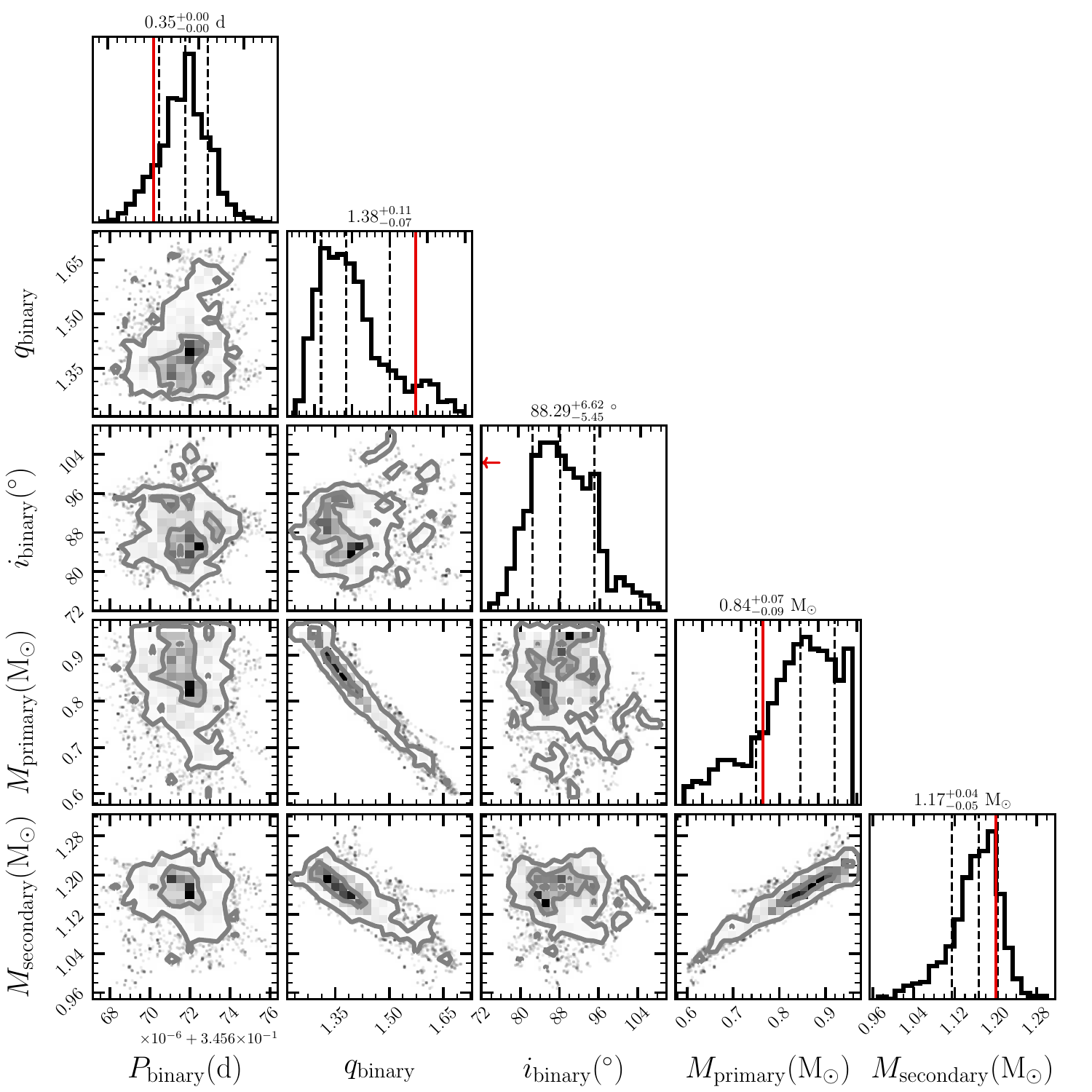}
    \caption{MCMC posteriors for the synthetic system. The red lines indicate the injected values. The synthetic binary inclination is outside the plot range, so an arrow is shown instead.}
    \label{fig:jtest_corner}
\end{figure*}

The \PHOEBE{} models of J1208 and J1721 produce different predictions of the binary masses and inclinations depending on the \TESS{} sector used and the number of spots. Here, we create a synthetic binary and attempt to recover the ``true'' parameters using the same \PHOEBE{} modeling process.

We generate a binary with a primary mass of $M_1=0.765\ M_\odot$ and a secondary mass of $M_2 = 1.2\ M_\odot$. We use a MIST evolutionary track with Solar abundances to select $T_{\rm{eff}}=4760$~K and $R_1=0.69\ R_\odot$ for the primary. The orbital inclination is set to $i=72^{\circ}$ and the orbital period is $P=0.34567$~days. We add a single spot to the primary star, with position $\theta_s = 47^{\circ}$, $\phi=82^{\circ}$, size $R_s=13^{\circ}$ and relative spot temperature $T_s/T_{\rm{eff}}=0.80$.

We created a synthetic light curve using times matching the \TESS{} observations from the J1721 and selected 5 random times for the RV observations. We then add random noise to each dataset with uncertainties on the normalized flux of $1\times10^{-4}$ and on the RVs of $0.1$~km/s. We follow the same \PHOEBE{} modeling steps described in Section \S\ref{sec:lcs}. Figure \ref{fig:jtest_corner} shows the posteriors as compared to the injected values. Although we find a secondary mass $M_2=1.17^{+0.04}_{-0.05}\ M_\odot$ consistent with the true mass, the orbital inclination and mass ratios are not recovered. 

The spot positions also differ from the injected values. The \PHOEBE{} model prefers a spot with a higher temperature ratio $T_s/T_{\rm{eff}}=0.95$ and a slightly larger size $R_s=21^{\circ}$. The spot position is also different with $\theta_s=83^{\circ}$ and $\phi_s = 85^{\circ}$. 

Although we only generated one synthetic binary, this test illustrates the challenges in recovering orbital parameters for spotted ellipsoidal variables. The companion mass was recovered correctly in this case, but the inclination and spot parameters differ from the true values, suggesting that it is still risky to trust the results.


\bsp	
\label{lastpage}
\end{document}